\documentclass[journal]{IEEEtran}

\usepackage[utf8]{inputenc}
\usepackage[margin=0.70in]{geometry}
\usepackage{mathtools}
\usepackage{amssymb}
\usepackage{algorithm}
\usepackage{algpseudocode}
\usepackage{comment}
\usepackage{lipsum}
\usepackage{graphicx}
\usepackage{subcaption}
\usepackage{caption}
\usepackage{multicol}
\usepackage{xcolor}
\usepackage{amsthm}
\usepackage{Symbol}
\usepackage{algcompatible}
\usepackage{siunitx}
\usepackage{booktabs}
\usepackage{color, colortbl}
\usepackage{enumitem}
\usepackage{cuted}
\usepackage{url}

\usepackage[noadjust]{cite}

\makeatletter
\newcommand*{\rom}[1]{\expandafter\@slowromancap\romannumeral #1@}

\newtheorem{proposition}{Proposition}

\newtheorem{remark}{Remark}

\newtheorem{assumption}{Assumption}
\newtheorem{corollary}{Corollary}
\newtheorem{lemma}{Lemma}

\begin{document}
\title{Graph signal aware decomposition of dynamic networks via latent graphs}
\author{{Bishwadeep~Das},~\IEEEmembership{Member,~IEEE,}~Andrei~Buciulea,~\IEEEmembership{Member,~IEEE,}~Antonio~G.~Marques,~\IEEEmembership{Senior~Member,~IEEE}~and~Elvin~Isufi,~\IEEEmembership{Senior Member,~IEEE}\thanks{This research is supported by the TTW-OTP project GraSPA (project number 19497) financed by the Dutch Research Council (NWO) and by the TU Delft AI programme. This work was partially supported by the the EU H2020 Grant Tailor (No 952215, agreements 31 and 82), the Spanish AEI (10.13039/501100011033) grants PID2022-136887NB-I00, TED2021-130347B-I00, and the Community of Madrid via the ELLIS Madrid Unit and the grant IDEA-CM (TEC-2024/COM-89). 
Das and Isufi are with the Faculty of Electrical Engineering, Mathematics and Computer Science, Delft University of Technology, The Netherlands, e-mails: \{b.das,e.isufi-1\}@tudelft.nl. Buciulea and Marques are with the Department of Signal Theory and Communications, King Juan Carlos University, Madrid, Spain. e-mails: \{andrei.buciulea,antonio.garcia.marques\}@urjc.es. A preliminary version of this work has been presented at ICASSP 2024 \cite{das2024tensor}.}}
\maketitle
\begin{abstract}
Dynamics on \textit{and} of networks refer to changes in topology and node-associated signals, respectively and are pervasive in many socio-technological systems, including social, biological, and infrastructure networks. Due to practical constraints, privacy concerns, or malfunctions, we often observe only a fraction of the topological evolution and associated signal, which not only hinders downstream tasks but also restricts our analysis of network evolution. Such aspects could be mitigated by moving our attention at the underlying latent driving factors of the network evolution, which can be naturally uncovered via low-rank tensor decomposition. Tensor-based methods provide a powerful means of uncovering the underlying factors of network evolution through low-rank decompositions. However, the extracted embeddings typically lack a relational structure and are obtained independently from the node signals. This disconnect reduces the interpretability of the embeddings and overlooks the coupling between topology and signals.
To address these limitations, we propose a novel two-way decomposition to represent a dynamic graph topology, where the structural evolution is captured by a linear combination of latent graph adjacency matrices reflecting the overall joint evolution of both the topology and the signal. Using spatio-temporal data, we estimate the latent adjacency matrices and their temporal scaling signatures via alternating minimization, and prove that our approach converges to a stationary point. Numerical results show that the proposed method recovers individually and collectively expressive latent graphs, outperforming both standard tensor-based decompositions and signal-based topology identification methods in reconstructing the missing network—especially when observations are limited.
\end{abstract}

\IEEEpeerreviewmaketitle
\section{Introduction}
Dynamic or temporal networks undergo structural evolution in the form of changing edges between nodes \cite{holme2012temporal,casteigts2012time,lindquist2009network,zhang2017random}. Dynamic networks co-occur alongside dynamic graph signals, which represent the evolution of a process over the nodes. Since graph signals depend on the supporting nodes and their connections, dynamic graph signals convey important information about topological changes. Identifying graphs from dynamic graph signals has been approached from several viewpoints, wherein the topology is usually inferred through a batch or online learning problem that seeks to minimize loss functions with graph-based priors \cite{kalofolias2017learning, natali2022learning,money2023sparse,shafipour2020online,shen2017tensor,baingana2016tracking,hallac2017network,buciulea2024online,javaheri2023joint}. One can also infer multiple graphs from dynamic graph signals. For example, in \cite{karaaslanli2024multiview}, the authors learn multi-view graphs that vary around a single consensus graph. The work in \cite{he2022online} learns multiple graphs from streaming node signals that differ in terms of their sets of central nodes, determined via eigen-centrality. The work in \cite{park2014graph} learns a set of graphs from observed spatiotemporal variations using spatial independent component analysis \cite{hyvarinen2000independent}. The notion of having a set of known or unknown underlying graphs has been used to perform signal processing tasks in \cite{pedersen2018multilayer,lu2020graph,baingana2016tracking,liu2023graph,davison2015brain}. These approaches infer multiple graphs solely from signals under various settings. A drawback of topology identification from graph signals (static or dynamic) is the choice of the prior, which may not be known in advance, may be misaligned with the data, or may require extensive validation.

\par Another choice for representing the structure of dynamic networks is by using tensors \cite{cichocki2016tensor}. This is typically done by stacking their topological matrix representation along the temporal dimension to obtain a three-dimensional tensor. One advantage of such a representation is that it allows for low-rank tensor decomposition \cite{kolda2009tensor,de2000multilinear,de2008decompositions}, which aids in compression and representation of the dynamic network, facilitating downstream tasks. 
These tensor-based decompositions of dynamic networks have resulted in various applications. For example, the works in \cite{gauvin2014detecting,gorovits2018larc,gujral2020beyond} use low-rank tensor decomposition for community detection. The work in \cite{fernandes2018dynamic} applies tensor decomposition for graph summarization. An example of link prediction can be seen in \cite{dunlavy2011temporal}, but the tensor here represents a changing heterogeneous user-item graph. Another example involves anomaly detection \cite{sapienza2015anomaly}. The works in \cite{mahyari2014identification,mahyari2016tensor,ozdemir2017recursive} build tensors from MRI data and use low-rank decomposition for tracking the states of dynamic networks. However, these tensor-based approaches face the following limitations:
\begin{itemize}
\item \textit{Role of the graph}. They treat the adjacency tensor purely from a tensor perspective, overlooking the graph structure it represents. Consequently, they often yield low-rank decompositions that lack specific graph-related information. This limitation arises partly because the focus is typically on downstream tasks rather than on capturing topological properties. Nonetheless, incorporating graph structure in the decomposition is crucial for uncovering hidden relationships in the embedding, which are typically enforced via priors. The work in \cite{qiu2020generalized} uses the tensor data to construct a graph for each dimension and use them as a regularizer to obtain latent embeddings. The embeddings themselves do not have any relational structure. Moreover, such approaches do not extend to tensors capturing time-varying graph structures.
\item \textit{Leveraging graph signals}. Existing tensor decompositions often overlook the relationship between the graph's topology and the node-associated signals (data). In many cases, graph signals are not considered at all. Even when they are considered, the data are frequently used as a proxy for the tensor to be learned. By contrast, leveraging graph signals to identify underlying topologies is crucial in several topology identification approaches \cite{mateos_connecting_2019,buciulea2022learning,navarro2024joint}, leading to more accurate topological representations. It can also uncover more meaningful latent graphs that jointly shape both topological and signal evolution.
\end{itemize}

\par To address some of these limitations, we draw inspiration from the multi-graph perspective of dynamic graphs and their relationship with graph signals, and propose a novel topology-and-signal-aware decomposition of partially observed dynamic networks. At its core, our decomposition relies on a set of latent graphs whose linear combination captures the observed structural evolution. Moreover, we exploit dynamic graph signals to facilitate the recovery of these latent graphs, particularly when the dynamic topology is only partially observed. Consequently, our approach recovers both the latent graphs and their associated temporal signatures that modulate their contributions over time. This framework is characterized by the following features:
\begin{enumerate}
\item It leads to interpretable latent graph decompositions.
\item It incorporates spatiotemporal nodal signals as priors to enhance the estimation of dynamic networks, especially when the topological observations are limited.
\item It uses of an alternating optimization algorithm to recover the latent graphs and their corresponding temporal signatures, with convergence guarantees to stationary points.
\item It demonstrates strong performance in reconstructing unobserved network structures, outperforming low-rank tensor decompositions and other topology-agnostic alternatives.
\end{enumerate}

\par The remainder of the paper is organized as follows. In Section \ref{J3 Section PF}, we introduce the dynamic network decomposition model and formulate our problem. In Section \ref{J3 Section method}, we present an alternating optimization algorithm to solve the dynamic network decomposition under the specified constraints. In Section \ref{J3 Section Analysis}, we analyze the convergence properties of the proposed algorithm. In Section \ref{J3 Section Experimental}, we conduct experiments to highlight various aspects of the decomposition and compare its performance with alternative approaches. Finally, Section \ref{J3 Section Conclusion} concludes the paper. All proofs are provided in the Appendix.

\par\noindent\textbf{Notation.} The Kronecker product between two matrices $\bbA$ and $\bbB$ is denoted as $\bbA\otimes\bbB$. The operation $\textnormal{vec}(\bbA)$ converts matrix $\bbA$ into a vector by stacking its columns. Its inverse, $\textnormal{vec}^{-1}(\bba)$, converts the vector $\bba$ to $\bbA$. The $M\times N$ matrices of ones and zeros are denoted as $\bbone_{M\times N}$ and $\bbzero_{M\times N}$, respectively.
\section{Problem Formulation}\label{J3 Section PF}
\begin{figure*}
    \centering
    {\includegraphics[trim=0 0 0 0,clip,width=0.8\textwidth]{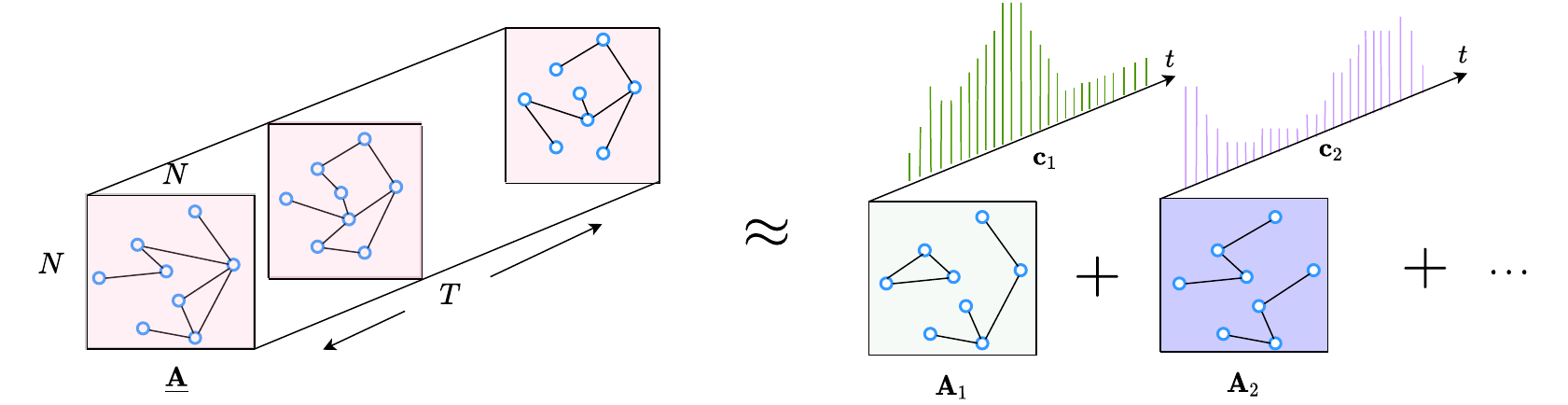}}%
\caption{Decomposition of the temporal topology evolution in terms of the component adjacency tensors. (left) an evolving topology over $N$ nodes for $T$ time instants stacked into a three-dimensional tensor $\uvbA$. This is expressed as a sum of latent adjacency matrices modulated by their temporal signatures. (right) we show a representation of two latent graph adjacency matrices $\bbA_1$ and $\bbA_2$ and their corresponding temporal signatures $\bbc_1$ and $\bbc_2$. The linear combination of these graphs approximate $\uvbA$.}
\label{general illustration}
\end{figure*}
Consider a dynamic undirected graph $\ccalG_t = (\ccalV, \ccalE_t)$ with a fixed node set $\ccalV = {1, \ldots, N}$ and a time-varying edge set $\ccalE_t$ evolving over $t=1,\ldots,T$. We represent its evolution via a three-dimensional adjacency tensor $\uvbA \in \reals^{N\times N\times T}$, where the $t$-th frontal slice $\uvbA_{:,:,t} \in \reals^{N\times N}$ is a symmetric adjacency matrix at time $t$. The entry $[\uvbA_{:,:,t}]_{ij}=[\uvbA_{:,:,t}]_{ji} > 0$ indicates an edge between nodes $i$ and $j$ at time $t$, and $[\uvbA_{:,:,t}]_{ij} = 0$ otherwise.

Frequently, we do not observe the complete topology at any given instant. This occurs, for example, in evolving social networks if some users do not reveal their connections for privacy reasons \cite{houghton2014privacy}, or in sensor networks that experience communication outages in harsh environments \cite{gungor2010opportunities}. We account for these missing observations using a three-dimensional binary tensor mask $\uvbM \in \{0,1\}^{N\times N\times T}$, where $[\uvbM_{:,:t}]_{ij} = [\uvbM_{:,:t}]_{ji} = 1$ only if we observe the presence or absence of an edge between nodes $i$ and $j$ at time $t$. As a result, we see only part of the structural evolution, described by $\uvbM \circ \uvbA$ with $\circ$ denoting the element-wise Hadamard product. Since observations often include errors, we typically have access to $\uvbM \circ (\uvbA + \uvbE)$, where $\uvbE$ represents an error tensor.

A common way to reduce the degrees of freedom in $\uvbA$ and estimate unobserved edges is via low-rank tensor completion. For instance, canonical polyadic or block-term decompositions with very low rank \cite{kolda2009tensor,de2008decompositions} make recovery tractable but yield temporal adjacency matrices that are often low rank, which may not align with many real-world graph structures where, for a single time instant, multiple network effects and configurations are taking place jointly \cite{mateos_connecting_2019}. In contrast, we propose modeling each time-$t$ adjacency matrix in $\uvbA$ as a weighted sum of $R \!\ll\! T$ latent deterministic topologies (i.e., adjacency matrices of size $N \times N$). These latent graphs represent static connections that account for hidden interactions among the nodes. Multiple latent graphs can operate concurrently if certain types of interactions persist over time, or they can switch on and off, as in switching dynamical models \cite{fox2008nonparametric,liu2023graph}. Altogether, these latent graphs collectively capture the observed network’s structural evolution, facilitate interpretability and, if eventually required, some (all) of them can be promoted to be low rank as well.

More specifically, let the $r$-th adjacency matrix be $\bbA_r\in\reals^{N\!\times\!N}$ and consider the vector $\bbc_r\in\reals^{T}$ with $[\bbc_r]_t$ denoting the scalar weight of $\bbA_r$ at time $t$. We model the time-varying network as
\begin{align}\label{decomp_model}
\uvbA=\sum_{r=1}^R\bbA_r\diamond\bbc_r.
\end{align}
In words, we model the $N\!\times \! N \!\times \! T$ adjacency tensor $\underline\bbA$ as the sum of outer products $\diamond$ between the adjacency matrices ${\bbA_r}$ and their corresponding temporal signature vectors ${\bbc_r}$. Clearly, if we focus on a particular time instant $t$, the adjacency matrix $\uvbA_{:,:,t}$ is
\begin{align}
\uvbA_{:,:,t}=\sum_{r=1}^R[\bbc_r]_t\bbA_r
\end{align}
i.e., a linear combination of the $R$ latent adjacency matrices, each scaled by the corresponding elements in $\bbc_r$. Figure \ref{general illustration} shows the adjacency tensor $\uvbA$, along with the corresponding model approximation in terms of latent adjacency matrices $[\bbA_1,\bbA_2,...]$ and their temporal signatures $[\bbc_1,\bbc_2,...]$. Similar to low-rank decompositions, the adjacency matrices can also reveal important hidden connectivity factors and be used for subsequent downstream tasks such as temporal link prediction, graph reconstruction, or anomaly detection \cite{dunlavy2011temporal,sapienza2015anomaly}. The elements of $\bbc_r$ are temporal functions that modulate the latent graphs.

Besides the topology, we often have access to time-varying signals at each node. To formalize this, let $\uvbX\in\mathbb R^{N\times Q\times T}$ be the signal tensor associated with the dynamic topology in $\uvbA$. The signal matrix at time $t$ is $\uvbX_{:,:,t} \in\reals^{N\!\times\!Q}$ and relates to the observed topology in $\uvbA_{:,:,t}$, i.e., a $Q$-dimensional feature vector associated with each node. Under the assumption that the evolution of signals is closely related to the underlying topology \cite{natali2022learning,kalofolias2017learning,shafipour2020online,money2023sparse}, we use the signals in $\uvbX$ to recover the latent adjacency matrices and their temporal signatures, especially when we have missing observations in the evolving topology in $\uvbA$. For notational convenience, we sometimes denote $\uvbA_{:,:t}$, $\uvbM_{:,:t}$, and $\uvbX_{:,:t}$ as $\bbA_t$, $\bbM_t$, and $\bbX_t$, respectively.
\smallskip
\par\noindent\textbf{Problem formulation.} Given the observation mask $\uvbM$, the observed evolution of the network topology $\uvbM\circ (\uvbA + \uvbE)$ and the spatiotemporal signals $\uvbX$, the goal is to recover the $R$ adjacency matrices $\bbA_1,\ldots,\bbA_R$ of the latent graphs along with their temporal signatures $\bbc_1,\ldots,\bbc_R$, respectively.
\par Let us define the matrices $\bbA$ and $\bbC$ 
\begin{equation}
\bbA=[\bbA_1,...,\bbA_R],~\bbC=[\bbc_1,...,\bbc_R].
\end{equation}which stack the matrices and their signatures horizontally.
We solve for these by addressing the following general optimization problem 
\begin{align}\label{eq.General}
&\underset{\{\bbA_r\},\{\bbc_r\}}{\text{minimize}}~
f(\bbA,\bbC)~=\ell(\uvbA,\uvbM,\bbA,\bbC)\!\!+\!\!\sum_{r=1}^R \pi_A(\bbA_r)\! +\! \pi_C(\bbC) \nonumber\\
&~\hspace{2.7cm}+g(\uvbX,\bbA,\bbC)+h(\bbA) \\
&\text{subject to}~\bbA_r\in\ccalS_A,~\bbA=[\bbA_1,\ldots,\bbA_R],\nonumber\\
&~\hspace{1.4cm}\bbc_r\in\ccalS_C,~\bbC=[\bbc_1,\ldots,\bbc_R].\nonumber
\end{align}
The function $\ell()$ is the fitting term, measuring how well the latent adjacency matrices $\bbA_r$s and their temporal signatures $\bbc_r$s approximate the observed part of $\uvbA$, functions $\pi_A()$ and $\pi_C()$ impose structural priors on each of the latent adjacency matrices $\bbA_r$s and the temporal signatures in $\bbC$, respectively. The function $g()$ accounts for the relationship between the signal tensor $\uvbX$ and the latent adjacency matrices. The function $h()$ is a prior on $\bbA$, or more specifically, the relationship across the different $\bbA_r$ (e.g., penalizing latent matrices with a similar support). The sets $\ccalS_A$ and $\ccalS_C$ impose strict structural constraints on the respective variables.
\par This formulation has a high number of degrees of freedom, with $N^2$ for each latent graph and $T$ for each temporal signature. This leads to a total of $N^2R+TR$ degrees of freedom. To make the problem tractable, we need to use structural constraints as priors, which we will detail in Section \ref{J3 Section method}. Moreover, when the number of observations in $\uvbM\circ\uvbA$ is smaller than $N^2R+TR$, the constraints and priors become even more relevant in the context of this under-determined problem.
\section{Dynamic Graph Decomposition}\label{J3 Section method}
\par Problem \eqref{eq.General} is generic and can be tailored to diverse contexts. In this section, we refine it by specifying the fitting error loss function, along with the constraints on the latent adjacency matrices in $\bbA$ and their temporal signatures in $\bbC$. We also define functions that promote the desired relationships among the latent adjacency matrices $\bbA$ and between the signals $\uvbX$, the adjacency matrices $\bbA$, and the temporal signatures in $\bbC$.
\par\smallskip\noindent\textbf{Fitting error.} For the fitting function, we consider a least-squares (LS) cost
\begin{equation}\label{eq.fittingterm_LS}
\ell(\uvbA,\uvbM,\bbA,\bbC)=||\uvbM\circ(\uvbA-\sum_{r=1}^R\bbA_r\diamond\bbc_r)||_F^2,
\end{equation}
that measures how well the latent graphs predict observed topological evolution. The norm in \eqref{eq.fittingterm_LS} can be adjusted to incorporate prior information about the observation errors. For example, the $\ell_1$ norm would be well-motivated if errors were sparse and independent across links. 

\par\smallskip\noindent\textbf{Constraint sets.} We want to estimate latent graphs that are undirected, have positive weights and do not contain self-loops. As a result, we consider $\ccalS_A = \{\bbS \in \reals^{N \times N}: \, \tr(\bbS) = 0,\, \bbS=\bbS^{\top}, \, [\bbS]_{ij} \geq 0 \, \forall \,(ij), \}$, representing non-negative symmetric $N\times N$ matrices with zero trace. For the temporal signatures, we consider $\ccalS_C=\{\bbs \in \reals^{T}:\,[\bbs]_i\geq 0 \;\forall\, i\}$. Positive weights in both $\bbA_r$ and $\bbc_r$ aid in interpretation, especially when it comes to comparing different time signatures. They also further restrict the degrees of freedom, rendering Problem \eqref{eq.General} more tractable.
\smallskip
\par\noindent\textbf{Sparsity of latent graphs.} We consider the latent graphs to be sparse. We impose sparsity on each adjacency matrix via $\pi_A(\bbA_r)=||\bbA_r||_1$. For $\bbA_r\in\ccalS_A$, this means $\pi_A(\bbA_r) = \mathbf{1}_N^{\top}\bbA_r\mathbf{1}_N$, where $\bbone_N$ is the column vector of $N$ ones.
\smallskip
\par\noindent\textbf{Signal smoothness.} We assume the graph signals are smooth on the (unobserved) graph \cite{dong2016learning,kalofolias2017learning}. Signal smoothness, which implies that nodes connected by edges have similar values, has been commonly used as a prior for identifying topology and performing inference over graphs \cite{mateos_connecting_2019}. To be more specific, we assume the signals at time $t$, $\bbX_t$, to be smooth over the approximated topology at the same time, given by $\sum_{r=1}^R[\bbC]_{t,r}\bbA_r$. A graph signal is smooth if nodes that share edges have similar signal values. Consider the matrix $\bbZ_t\in\mathbb{R}^{N\times N}$ with its $(ij)$-th element $[\bbZ_t]_{ij}=||[\bbX_t]_{i,:}-[\bbX_t]_{j,:}||_2^2$, measuring the squared $\ell_2$ norm between the signals $[\bbX_t]_{i,:}$ and $[\bbX_t]_{j,:}$ at nodes $i$ and $j$ at time $t$. The smoothness of the graph signal $\bbX_t$ over the recovered model-driven topology is given by the quadratic variation $\QV(\bbX_t)=\frac{1}{2}\tr(\uvbA_{:,:,t}\bbZ_t)$ \cite{kalofolias2017learning}. The signal-incorporating prior $g()$ is the sum of the signal smoothness function for each time instant, and it is calculated as
\begin{align}\label{prior smoothness}
\begin{split}
g(\uvbX,\bbA,\bbC)&=\sum_{t=1}^T\frac{1}{2}\tr(\bbA_{:,:,t}\bbZ_t)=\sum_{t=1}^T\sum_{r=1}^R[\bbC]_{t,r}\frac{\tr(\bbA_r\bbZ_t)}{2}.
    \end{split}
\end{align}
This promotes the creation of new links from the signals in $\uvbX$ that are smooth in the original temporal network when portions of $\uvbA$ are not observed due to the mask. Finally, notice that we do not impose temporal smoothness. However, our model can accommodate it by a straightforward modification of the cost function in \eqref{prior smoothness}.
\smallskip
\par\noindent\textbf{Non-overlapping support.} We desire that the supports of the different latent graphs do not overlap so that each component uniquely contributes to the structural evolution and provides distinct insights. One way to promote this non-overlapping support is to penalize the $\bbA_r$s that share many common edges. This is achieved by imposing a regularizer on the latent adjacency matrices through the function
\begin{align}
h(\bbA)=\sum_{r=1}^R\sum_{\substack{\bar{r}=1 \ \bar{r}\neq r}}^R\tr(\bbA_r^{\top}\bbA_{\bar{r}}),
\end{align}
which promotes orthogonality among the latent adjacency matrices. The lowest possible value of $h(\bbA)$ is zero, which is achieved when no edge is shared among the adjacency matrices.
\smallskip
\par\noindent\textbf{Avoiding trivial solutions.} The formulation in \eqref{eq.General} can generate trivial solutions for some of the variables. More specifically, suppose we have a mask $\uvbM$ that hides all the information across an edge for all $t$, i.e., $[\uvbM_{:,:,t}]_{ij}=[\uvbM_{:,:,t}]_{ji}=0~\forall~t$. This leads to a trivial solution for the corresponding variables because of the dependence only on the functions $g()$ and $h()$. One way to tackle this is to force each node at every time instant to have a positive degree $\zeta>0$. This is expressed through the constraint
\begin{align}\label{constraint for nontrivial}
\Big(\sum_{r=1}^R [\bbC]_{t,r}\bbA_r\Big)\mathbf{1}_N \geq \zeta \mathbf{1}_N.
\end{align}
We extend this constraint for all time instants as
\begin{align}
\bbA[\bbC^{\top}\otimes\mathbf{1}_N]\geq \zeta\mathbf{1}_{N\times T},
\end{align}
where we recall that $\bbA=[\bbA_1,\ldots,\bbA_R]$. This implies that each node in the reconstructed topology at time $t$ has a degree of at least $\zeta$.
\smallskip
\par\noindent\textbf{Smooth temporal signatures.} We consider each temporal signature $\bbc_r$ to change slowly over time. This helps prevent overfitting to the observed topology $\uvbM\circ\uvbA$, especially when fewer observations are present. A gradual change in the temporal factor leads to smaller changes in the latent adjacency matrices.
A gradual change in the temporal signature also helps in predicting an edge if it is not observed over an interval of time, given the true structural evolution is slow. We use  the squared Frobenius norm for the matrix formed by the temporal difference of the elements of each $\bbc_r$, i.e,
\begin{align}
\begin{split}
\pi_C(\bbC)=||\bbD\bbC||_F^2=&\sum_{r=1}^R\sum_{t=2}^T([\bbC]_{t,r}-[\bbC]_{t-1,r})^2\\
\end{split}
\end{align}
where $\bbD$ is the $T\!\times\!(T\!-\!1\!)$ temporal difference matrix. Additionally, we consider an extra cost function for $\bbC$, namely the squared Frobenius norm. These functions are reasonable from a practical point of view and aid in the convergence analysis of the optimization as we shall elaborate in Section \ref{J3 Section Analysis}.
\par With all these in place, the optimization in \eqref{eq.General} can now be written as
\begin{align}\label{Opt prob basic}
    \begin{split} &\underset{\{\bbA_r\},\{\bbc_r\}}{\text{minimize}}~~f(\bbA,\bbC)\!=\!\frac{1}{2}\Big\|\uvbM\circ(\uvbA-\sum_{r=1}^R\bbA_r\diamond\bbc_r)\Big\|_F^2\!\!\\&+\gamma\!\sum_{r=1}^R\mathbf{1}_N^{\top}\bbA_r\mathbf{1}\! +\!\delta\sum_{t=1}^T\sum_{r=1}^R[\bbC]_{t,r}\frac{\tr(\bbA_r\bbZ_t)}{2}\!+\! \mu||\bbD\bbC||_F^2\\&+\!\!\beta\sum_{r=1}^R\sum_{\bar{r}=1,\bar{r}\neq r}^R\tr(\bbA_r^{\top}\bbA_{\bar{r}})+\frac{\rho}{2}||\bbC||_F^2\\
    &\text{subject to}\\&\bbA_r\in\ccalS_A = \{\bbA_r:\bbA_r \geq \bbzero_{N\times N}, \tr(\bbA_r) = 0 ,\bbA_r=\bbA_r^{\top}\},
\\&\bbc_r\in\ccalS_C=\{\bbc_r:\bbc_r\geq\mathbf{0}_{T}\},\\&[\bbA_1,\ldots,\bbA_R][\bbC^{\top}\otimes\mathbf{1}_N]\geq \zeta\mathbf{1}_{N\times T},
    \end{split}
\end{align}
where $[\gamma,\delta,\beta,\mu, \rho]^{\top}$ denote the hyper-parameters.
\begin{remark}
Problem \eqref{Opt prob basic} is a particular instance of the more general Problem \eqref{eq.General}. The constraints and regularizers can be modified to model diverse scenarios. For example, low-rank latent factors can be promoted via the use of nuclear norms. Additionally, a switching dynamic network can be captured by penalizing the $\ell_1$ norm of each column in $\bbC$ (ensuring that most latent graphs remain inactive at each time instant), the operator $\bbD\bbC$ (preventing the latent graph from changing too frequently), or combinations of the two.
\end{remark}

\subsection{Solving the decomposition}
Due to the multiple bilinear terms, Problem \eqref{Opt prob basic} is jointly non-convex in $\{\bbA_1,...,\bbA_R,\bbC\}$. However, it is separately convex in each variable when the others are fixed. This makes it easier to solve for one variable at a time via an alternating approach, i.e., we solve a sequence of convex problems in these variables until convergence. We do this by formulating $R+1$ optimization problems, where each of the first $R$ solves for a latent adjacency matrix $\bbA_r$, whereas the last solves for the temporal signatures in $\bbC$. We solve each of them via the alternating direction method of multipliers. Thus, we perform alternating minimization between $R+1$ variables.
\subsection{Updating the $\bbA_r$s}
Here we outline the update for the $r$-th latent adjacency matrix $\bbA_r$s, keeping fixed the other matrices $\{\bbA_{r'}\}_{r'\neq r}$ and the temporal signatures in $\bbC$. Consider the $N^2\times 1$ vector $\bbm_t$ formed by vectorizing the mask at time $t$ $\bbM_t$, i.e., $\bbm_t=\text{vec}(\bbM_t)$. Consider also the $N^2\times T$ matrix $\bbM_0$ that collects the $\bbm_t$s along its columns, i.e., \begin{equation}\label{eq M_0}
    \mathbf{M_{0}}=[\bbm_1,\bbm_2,\ldots,\bbm_T]\in\mathbb{R}^{N^2\times T}.
\end{equation}
Let us define the matrix $\bbXi_r=\frac{1}{2}[\bbZ_1,\ldots,\bbZ_T](\bbc_r\otimes\bbI_N)$; then, the terms in the loss function $f(\cdot)$ that depend on the $r$-th adjacency matrix $\bbA_r$ can be rewritten as 
%
\begin{align}\label{J3 loss in A_r}
    \begin{split}
 & f(\bbA_r)=\frac{1}{2}||\bbA_r||_F^2\sum_{t=1}^T[\bbC]_{t,r}^2\mathbf{1}_{N^2}^{\top}\bbm_t \\&+\tr\bigg(\bbA_r^{\top}\big[\sum_{t=1}^T[\bbC]_{t,r}\mathbf{1}_{N^2}^{\top}\bbm_t\sum_{\bar{r}=1,\bar{r}\neq r}^R[\bbC]_{t,\bar{r}}\bbA_{\bar{r}}\big]\bigg)\\&-\tr\big(\bbA_r^{\top}\text{vec}^{-1}\big(\sum_{t=1}^T\uvbA_{:,:,t}[\bbC]_{t,r}\mathbf{1}_{N^2}^{\top}\bbm_t\big)\big)+\gamma\mathbf 1_N^{\top}\bbA_r\mathbf 1_N\!\\&+\!\delta\tr(\bbA_r\bbXi_r)+2\beta\text{vec}(\bbA_r)^{\top}\!\!\sum_{\substack{\bar{r}=1,\bar{r}\neq r}}^R\!\!\text{vec}(\bbA_{\bar{r}})\!\!\!.\end{split}
\end{align} 
Note the term $\sum_{t=1}^T[\bbC]_{t,r}^2\mathbf{1}_{N^2}^{\top}\bbm_t$ that scales the squared Frobenius norm of $\bbA_r$. It is the sum across time of the product between the corresponding squared temporal coefficient $[\bbC]_{t,r}$ and the number of ones in the mask $\bbone^{\top}\bbm_t$, which is also the number of observations available in $\uvbA$ at that time. Note also that the influence of the mask on updating the $r$-th adjacency matrix is affected by the corresponding temporal signature $\bbc_r$. Upon defining the matrices $\bbPhi_r=\bbc_{r}^{\top}\otimes \mathbf1_N$ and $\bbGamma_r=\sum_{\bar{r}, \bar{r}\neq r}\bbA_{\bar{r}}(\bbc_{\bar{r}}^{\top}\otimes \mathbf1_N)-\zeta\mathbf1_{N\times T}$, the associated optimization can be compactly written as
\begin{align}\label{problem_Ar}
    \begin{split} &\underset{\bbA_r}{\text{minimize}}~f(\bbA_r)\\
    &\text{subject to}~\bbA_r\bbPhi_r+\bbGamma_r\geq\mathbf0_{N\times T},\\&\bbA_r \in \ccalS_A= \{\bbS:\bbS \geq \bbzero_{N\times N}, \tr(\bbS) = 0 ,\bbS=\bbS^{\top}\},
    \end{split}
\end{align}
where the inequality constraint is the same as in \eqref{constraint for nontrivial} but written in terms of $\bbA_r$. We solve Problem \eqref{problem_Ar} via the Alternating Direction of Method of Multipliers (ADMM) \cite{boyd2011distributed}. The formulation is
\begin{align}\label{augmented prob}
    \begin{split} &\underset{\bbA_r,\bbP}{\text{minimize}}~f(\bbA_r)\\
    &\text{subject to:}~\bbA_r \in \ccalS_A,~\bbA_r\bbPhi_r+\bbGamma_r=\bbP,~\bbP\geq\mathbf{0}_{N\times T},
    \end{split}
\end{align}
where the auxiliary variable $\bbP$ is introduced to cast the problem in the ADMM format. Then, the augmented Lagrangian of \eqref{augmented prob} is
\begin{align}\label{LagrangianAr}
    \begin{split}
L(\bbA_r,\bbP,\bbLambda_r)&=f(\bbA_r)+\tr(\bbLambda_r(\bbA_r\bbPhi_r+\bbGamma_r-\bbP))\\&+\frac{\lambda}{2}||\bbA_r\bbPhi_r+\bbGamma_r-\bbP||_F^2,
    \end{split}    
\end{align}
where $\lambda>0$. The updates in the primal variables $\bbA_r$ at iteration $k$ are given as 
\begin{align}\label{ADMM_A}
\begin{split}
    \bbA_r^{k} & = \ccalP_{\ccalS_A}\left(\bbA_r^{k-1}-\lambda\nabla_{\bbA_r}L(\bbA_r^{k-1})\right),
    \end{split}
\end{align}
which is a gradient step followed by a projection onto the set $\ccalS_A$. 
The expression for the gradient $\nabla_{\bbA_r}L(\bbA_r)$ is detailed in Appendix \ref{J3 Appendix ADMM Updates} [cf. \eqref{Appendix derivative A_r}]. For the auxiliary variable $\bbP$, we note that the Lagrangian is convex w.r.t. $\bbP$ and we can readily obtain a closed-form solution, followed by a projection onto the non-negative orthant. The remaining updates for the primal variable $\bbP$ and the dual variables $\bbLambda_r$ are thus
\begin{align}\label{ADMM_A_P}
    \bbP^{k}=[\frac{1}{\lambda}\bbLambda_r^{k-1\top}+\bbA_r^{k}\bbPhi_r+\bbGamma_r]_+
\end{align}
\begin{equation}\label{ADMM_A_Dual}
    \bbLambda_r^{k}=\bbLambda_r^{k-1} + \lambda(\bbA_r^{k}\bbPhi_r+\bbGamma_r-\bbP^{k})^\top,
\end{equation}
where $[\cdot]_+$ is an element-wise operator that clips all negative values to zero. The two steps are repeated until a stopping criterion is met. 
\subsection{Updating $\bbC$}
In this step, we update $\bbC$ keeping $\{\bbA_r\}_{r=1}^R$ fixed. As we did with the adjacency matrices, we first write down the loss function w.r.t. $\bbC$. To do so, we define 
\begin{equation}\label{eq X_0}
    \bbA_{\textnormal{vec}}=[\text{vec}(\uvbA_{:,:,1}),\text{vec}(\uvbA_{:,:,2}),\ldots,\text{vec}(\uvbA_{:,:,T})]\in\mathbb{R}^{N^2\times T}.
\end{equation}
\begin{equation}\label{J3 Define A0}
    \mathbf{A_{0}}=[\text{vec}(\bbA_1),\text{vec}(\bbA_2),\ldots,\text{vec}(\bbA_R)]\in\mathbb{R}^{N^2\times R}.
\end{equation}
The loss in $\bbC$ then writes as
\begin{align}
\begin{split}
    f(\bbC)&=\frac{1}{2}||\bbM_0\circ( \bbA_{\textnormal{vec}}-\bbA_0\bbC^{\top})||_F^2\! +\!\delta\tr(\bbC^{\top}\bar{\bbZ}\bbA_0)\\&+\frac{\mu}{2}||\bbD\bbC||_F^2+\frac{\rho}{2}||\bbC||_F^2,\\
\end{split}
\end{align}
where $\tr(\bbC^{\top}\bar{\bbZ}\bbA_0)$ is the signal smoothness prior \eqref{prior smoothness} re-written to show the dependence on  $\bbC$, and $\bar{\bbZ}\in\mathbb{R}^{T\times N^2}$ is a matrix with row $t$ being the vectorized form of the matrix $\bbZ_t^{\top}$, i.e., $[\bar{\bbZ}]_{:,t}=\text{vec}(\bbZ_t^{\top})^{\top}$. The inequality constraint involving both $\bbA$ and $\bbC$ in Problem \eqref{Opt prob basic} translates into $\bbC\Upsilon^{\top}-\zeta\mathbf1_{T\times N}\geq\mathbf0_{T\times N}$ where $\Upsilon=[\bbA_1,\ldots,\bbA_R](\bbI_R\otimes\mathbf1_N)$. Thus, the optimization problem for $\bbC$ can be compactly written as
\begin{align}\label{ProblemC}
    \begin{split} \underset{\bbC}{\text{minimize}}\quad &f(\bbC)\\\text{subject to: }&\bbC\Upsilon^{\top}-\zeta\mathbf1_{T\times N}\geq\mathbf0_{T\times N},\\&\bbC\in\ccalS_C=\{\bbS:\bbS\geq\mathbf0_{T\times R}\}.
    \end{split}
\end{align}
To deal with the inequality constraint, we formulate another ADMM problem as
\begin{align}
    \begin{split} \underset{\bbC}{\text{minimize}}\quad &f(\bbC)
    \\\text{subject to: }&\bbC\Upsilon^{\top}-\zeta\mathbf1_{T\times N}\!=\!\bbQ,\\&\bbC\in\ccalS_C, \;\bbQ\geq\mathbf0_{T\times N}.
    \end{split}
\end{align}
%
Then, the augmented Lagrangian becomes
\begin{align}\label{LagrangianC}
    \begin{split}
L(\bbC,\bbQ,\bbLambda)=& f(\bbC)+\tr\left(\bbLambda(\bbC\Upsilon^{\top}-\zeta\mathbf1_{T\times N}-\bbQ)\right)\\&+\frac{\lambda_c}{2}||\bbC\Upsilon^{\top}-\zeta\mathbf1_{T\times N}-\bbQ||_F^2.
    \end{split}
\end{align}To update $\bbC$, we take a gradient step followed by a projection onto $\ccalS_C$, i.e., 
\begin{align}\label{ADMM_C}
\begin{split}
    \bbC^{k} & = \ccalP_{\ccalS_A}\left(\bbC^{k-1}-\lambda\nabla_{\bbC}L(\bbC^{k-1})\right).
    \end{split}
\end{align} The derivative of the augmented Lagrangian w.r.t. $\bbC$ is detailed in Equation \eqref{Appendix derivative C} in Appendix \ref{J3 Appendix ADMM Updates}. The other two updates are also similar to those made for each $\bbA_r$ and are respectively
\begin{align}\label{ADMM_C_Q}
    \bbQ^{k}=[\frac{1}{\lambda_c}\bbLambda^{k-1\top}+(\bbC^{k}\bbUpsilon)^{\top}-\zeta\mathbf{1}_{TN})\big]_+,
\end{align}
\begin{align}\label{ADMM_C_Dual}
\bbLambda^{k}=\bbLambda^{k-1}+\lambda_c(\bbC^{k}\Upsilon^{\top}-\zeta\mathbf{1}_{TN}-\bbQ^{k})^\top.
\end{align}
Algorithm~\ref{Algorithm TGD}, termed Dynamic Graph Decomposition (DGD), summarizes the steps described in this section.
\begin{algorithm}[!t]
	\caption{\textbf{Dynamic Graph Decomposition (DGD)}}
	\begin{algorithmic}[1]
	\STATE \textbf{Input:} Observations $\uvbM\circ \uvbA\in\reals^{N\times N\times T}$ and $\uvbX\in\reals^{N\times Q\times T}$; mask $\uvbM$; parameters $R$, $\gamma$, $\beta$, $\mu$, $\delta$, $K$, $I$
    \STATE \textbf{Output:} Estimated latent graph adjacency matrices $\bbA_r=\bbA_r^{(I)}$ for $r=1,...,R$ and temporal signatures $\bbC=\bbC^{(I)}$
	\STATE \textbf{Initialization}: Initialize $\{\bbA_r^{(0)}\}_{r=1}^R$, $\bbC^{(0)}$ as i.i.d. realizations of a uniform random variable over $[0,1]$\\
    \quad\quad\textbf{For} $i=1:I$\\
    \quad\quad~~\textbf{$\bbA_r$-updates:} ($r=1,...,R$)\\
    \quad\quad\quad Initialize $\bbP^0$, $\bbLambda^0$ as i.i.d. realizations of  $\ccalN(0,1)$ \\    
    \quad\quad\quad Set $\bbC^0\!=\!\bbC^{(i-1)}$.\\
    \quad\quad\quad Set $\{\bbA_{r'}^0\!=\!\bbA_{r'}^{(i)}\}_{r'=1}^r$, $\{\bbA_{r'}\!=\!\bbA_{r'}^{(i-1)}\}_{r'=r+1}^R$\\
    \quad\quad\quad Form matrices $\bbXi_r$, $\bbPhi_r$, and $\bbGamma_r$    
    \\
    
    \quad\quad\quad\quad\textbf{For} $k=1:K$ (ADMM for  $\bbA_r$)\\
   \quad\quad\quad\quad\quad Update $\bbA_r^k$ using~\eqref{ADMM_A}\\
   \quad\quad\quad\quad\quad Update $\bbP^k$ using~\eqref{ADMM_A_P}\\
    \quad\quad\quad\quad\quad Update $\bbLambda^k$ using~\eqref{ADMM_A_Dual}\\
    \quad\quad\quad\quad\textbf{end For}\\
    \quad\quad\quad\quad$\bbA_{r}^{(i)} = \bbA_{r}^{K}$\\
\quad\quad~~\textbf{$\bbC$-update:}\\
\quad\quad\quad\quad Initialize $\bbQ^0$, $\bbLambda^0$ as i.i.d. realizations of $\ccalN(0,1)$\\
\quad\quad\quad\quad Set $\{\bbA_{r'}^0\!=\!\bbA_{r'}^{(i)}\}_{r'=1}^R$\\
    \quad\quad\quad\quad Form matrix $\Upsilon$
    \\
    \quad\quad\quad\quad\textbf{For} $k=1:K$ (ADMM for $\bbC$)\\
\quad\quad\quad\quad\quad Update $\bbC^k$ using \eqref{ADMM_C}\\
\quad\quad\quad\quad\quad Update $\bbQ^k$ using~\eqref{ADMM_C_Q}\\
\quad\quad\quad\quad\quad Update $\bbLambda^k$ using~\eqref{ADMM_C_Dual}\\
\quad\quad\quad\quad\textbf{end For}\\
\quad\quad\quad\quad $\bbC^{(i)} = \bbC^{K}$ \\
\quad\quad\textbf{end For}\\
\end{algorithmic} 
\label{Algorithm TGD}
\end{algorithm}
\section{Complexity and Convergence Analysis}\label{J3 Section Analysis}

In this section, we analyze the proposed DGD algorithm first in terms of its computational complexity and then in terms of the convergence behavior of the alternating approach.
\par\smallskip\noindent\textbf{Parameter complexity.} The parameters to be estimated are $\left(\frac{N-1}{2} + 2T\right)NR + (R+2N)T$. These come from two sources: (i) $\left(\frac{N-1}{2} + 2T\right)NR$ parameters associated with the $R$ matrices $\bbA_r$, which are symmetric with zero diagonals, and the remaining parameters are related to the additional variables $\bbP$ and $\bbLambda_r$; and (ii) $(R+2N)T$ parameters of the $R$ temporal signatures $\bbC$ and the additional variables $\bbQ$ and $\bbLambda$.
\par The number of parameters of the original problem [cf.\eqref{eq.General}] is $N^2T+TR$ 
\par\smallskip\noindent\textbf{Computational complexity.} Recalling that $I$ is the number of alternating iterations, the computational complexity of DGD is of order $\ccalO((N^3R + N^2TR + R^2N^2+ T^2R + NTR)I)$. For each iteration $i=1,...,I$, the computational complexity is broken down as follows. The computations for estimating $\bbA_r$ are governed by the gradient $\nabla_{\bbA_r}L(\bbA_r)$ in \eqref{ADMM_A} and the updates of the auxiliary variables $\bbP$ in \eqref{ADMM_A_P} and $\bbLambda_r$ in \eqref{ADMM_A_Dual}. We update the $R$ matrices $\bbA_r$, $\bbP$, and $\bbLambda_r$, leading to a total computational cost of $\ccalO((N^3 + N^2T)R)$. In addition, computing $\bbC$ has a cost of order $\ccalO((R^2T^2 + T^2R))$, and each update of the auxiliary variables $\bbQ$ and $\bbLambda$ has a complexity of order $\ccalO(NTR)$.
\begin{remark}
While the updates for $\bbA_r$ and $\bbC$ in \eqref{ADMM_A} and \eqref{ADMM_C} are carried out via projected gradient descent, one could alternatively derive a closed-form solution from the unconstrained problem and then apply a projection. However, this closed-form solution involves an inverse with complexity of order $\ccalO(N^3)$ for each $\bbA_r$ per ADMM iteration and $\ccalO(R^3T^3)$ for each $\bbC$ update since vectorization is required to obtain the optimal value. Moreover, projecting onto $\ccalS_A$ and $\ccalS_C$ afterward may introduce large adjustments, potentially affecting the convergence of Algorithm \ref{Algorithm TGD}. By contrast, a gradient step before projection is computationally less demanding, enforces more gradual change (due to the convex nature of the functions), and works effectively in practice.
\end{remark}
\subsection{Convergence analysis}\label{J3 Analysis}
In this section, we discuss the convergence of the proposed ADMM-based approach in Algorithm \ref{Algorithm TGD} to a stationary point w.r.t. the $\bbA_r$s and $\bbC$.
\begin{assumption}\label{J3 Assumption 1}
For all $r$, $\sum_{t=1}^T[\bbC]_{t,r}^2\bbone^{\top}\bbm_t>0$, i.e., the sum of the product of the squared temporal signature coefficient with the number of available observations across time is greater than zero.
\end{assumption}
This condition holds as long as the temporal signature of the $r$-th latent graph $\bbc_r\neq\bbzero_T$ across time and we have at least one available observation at each time instant. In practice, we make several observations at each time step, i.e., $\bbone^{\top}\bbm_t>0$ is easy to satisfy. A nonzero temporal signature $\bbc_r$ indicates that the corresponding latent adjacency matrix $\bbA_r$ is relevant to expressing the structural evolution.
\smallskip
\par\noindent\textbf{Convergence.} Since our decomposition approach is a block-wise minimization, we consider the block-coordinate minimizer as shown in \cite{buciulea2024polynomial,xu2013block}. Let $\chi$ be the set comprising all the latent variables $\{\bbA_1,\ldots,\bbA_R,\bbC\}$. The variables $\{\barbA_1,\ldots,\barbA_r,\barbC\}$ are a block coordinate minimizer of $f(\cdot)$ if we have
\begin{align}
&f(\barbA_1,\ldots,\barbA_r,\ldots,\barbA_R,\barbC)\leq f(\barbA_1,\ldots,\bbA_r,\ldots,\barbA_R,\barbC) \nonumber \\
&~\forall~r,\bbA_r\!\in\!\ccalS_{\bbA}\,\text{and}\,\{\barbA_1,\ldots,\bbA_r,\ldots,\barbA_R,\barbC\}\!\in\!\chi,\\
&f(\barbA_1,\ldots,\barbA_r,\ldots,\barbA_R,\barbC)\leq f(\barbA_1,\ldots,\barbA_r,\ldots,\barbA_R,\bbC)\nonumber\\
&~\forall~\bbC\in\ccalS_{\bbC}~\textnormal{and}~\{\barbA_1,\ldots,\barbA_R,\barbC\}\in\chi
\end{align}
\begin{proposition}\label{J3 Proposition Stationary Convergence}
    Let Assumption \ref{J3 Assumption 1} hold. Then, the sequence $\{\bbA_1^{k},\ldots,\bbA_R^k,\bbC^k\}_{i=1}^I$ generated by Algorithm \ref{Algorithm TGD}, as $K\rightarrow \infty$, converges to a stationary point, which is also known as the block coordinate minimizer of Problem \eqref{Opt prob basic}.
\end{proposition}
\par\noindent\textit{Proof.} See Appendix \ref{app:sectionB}.\qed
\section{Numerical Results}\label{J3 Section Experimental}
In this section, we validate the DGD algorithm by running experiments on: i) a synthetic controlled setting and ii) multiple real datasets. We first describe the datasets, evaluation metrics, and experimental details. Next, we analyze the properties of the proposed DGD method through a series of experiments. Finally, we compare it with alternative methods, focusing on the reconstruction error for the temporal network over a range of partial observations.
\subsection{Experimental setup}
We consider a synthetic experiment on stochastic block model graphs and three real datasets. The details of these datasets are summarized in Table \ref{details_datasets}.

\vspace{0.05cm}
\noindent \textbf{(1) Synthetic dynamic network (SwDyn).} We create a dynamic network that gradually transitions from a stochastic block model (SBM) graph with 2 communities to an SBM graph with 4 communities. The observed adjacency matrix at time $t$ is a linear combination of the two SBM graphs. Thus, we have $R=2$ latent graphs. We consider the temporal signature $\bbc_1$ of $\bbA_1$ as a vector with values decreasing linearly from one to zero over $50$ time instances. As a result, $\bbc_2=\bbone_{50}-\bbc_1$ with $\bbone_{50}$ the $50$ dimensional vector of ones has values increasing from zero to one, contributing to the emergence of $\bbA_2$. For each time step $t$, the signal feature matrix $\bbX_t$ is computed from the Laplacian matrix of $\uvbA_{:,:,t}$. More specifically, the (smooth) signals $\bbX_t$ are obtained by solving a Tikhonov-regularized least squares problem, where white noise is filtered using a Laplacian-based low-pass graph filter (see our code repository and \cite[Sec. 3]{kalofolias2016learn} for additional details). We generate a total of $Q = 1,000$ smooth graph signals associated with each one of the temporal graphs.


\begin{table}
\centering
\begin{tabular}{l l l l l l l}
dataset & $N$ & Av. E. &  $Q$ & $T$ & T-Var.(Std) & \QV\\
\hline
\textbf{SwDyn}      & 40   & 130   & 1000   & 50      & 0.17(0.08) & 0.08 \\
\textbf{SeaSurf} & 100   & 400   & 108    & 8      & 0.12(0.02) & 2.59 \\
\textbf{USTemp}   & 109   & 545   & 291     & 15      & 0.47(0.09) & 13.32 \\
\textbf{Contact} & 113   & 157  & 416   & 50      & 0.12(0.21) & 2.94
\\
\hline
\end{tabular}
\caption{Properties of all datasets. The columns represent the following: Number of nodes ($N$), average number of edges (Av. E.) per graph in the temporal network, number of nodal samples ($Q$), number of temporal graphs ($T$), temporal variation measured as the normalized squared Frobenius norm (T-Var.) between consecutive adjacency matrices along with the standard deviation (Std), and the normalized total quadratic variation for each node obtained from the signals (\QV).}
\label{details_datasets}
\vspace{-0.5cm}
\end{table}
\vspace{0.05cm}
\noindent \textbf{(2) Sea surface temperature (SeaSurf).} 
We consider surface temperature measurements across points on the Pacific Ocean \cite{giraldo2022reconstruction}. It consists of $1,728$ temporal measurements spread over $600$ months at $N=100$ points on the ocean. The temporal horizon was divided into $T=8$ equally sized windows. Since there is no ground-truth graph, we estimate one for each time window. This estimation is based on half of the available temporal measurements (those corresponding to odd time instants, $\tau={1,3,5,...}$), selecting the top $4N$ edges where the associated signal is smooth. The remaining half of the temporal measurements (those from even time instants, $\tau={2,4,6,...}$) are used as input for the numerical experiments, resulting in $Q = 108$ nodal measurements for each temporal graph.

\vspace{0.05cm}
\noindent \textbf{(3) US temperature (USTemp).} We use the NOAA dataset from \cite{arguez2012noaa}, which consists of 8,730 temporal graph signals indicating the temperature variation across $N=109$ stations. We divide the time into $T=15$ equally sized windows. Since a ground-truth graph is not available, we generate one for each window following similar steps to the SeaSurf dataset. Specifically, we estimate the graph using half of the temporal measurements (those corresponding to odd time instants, $\tau={1,3,5,...}$), selecting the top $5N$ edges where the associated signal is smooth. The remaining half of the temporal measurements (those corresponding to even time instants, $\tau={2,4,6,...}$) are used as input for the numerical experiments, resulting in $Q=291$ nodal measurements for each temporal graph.

\vspace{0.05cm}
\noindent \textbf{4) Contact.} This dataset was collected during the ACM Hypertext 2009 conference, where nodes represent $N=113$ attendees, and edges connect them based on face-to-face proximity.\footnote{http://www.sociopatterns.org/datasets/hypertext-2009-dynamic-contact-network/} The dynamic network captures interactions between attendees over two and a half days, resulting in a total of 20,800 interactions.
The signal at each time interaction is the sum of all prior interactions up to that point, normalized by the maximum number of interactions observed until that time. After constructing the network signals, we divide them into $T=50$ time windows, yielding $Q=416$ signals for each temporal graph. Since there is no ground-truth graph, we construct one for each time window by considering all interactions up to that point and selecting the top $3N$ edges corresponding to the nodes with the most interactions.

\begin{figure}
\centering
\includegraphics[trim=100 250 100 300,width=0.23\textwidth]{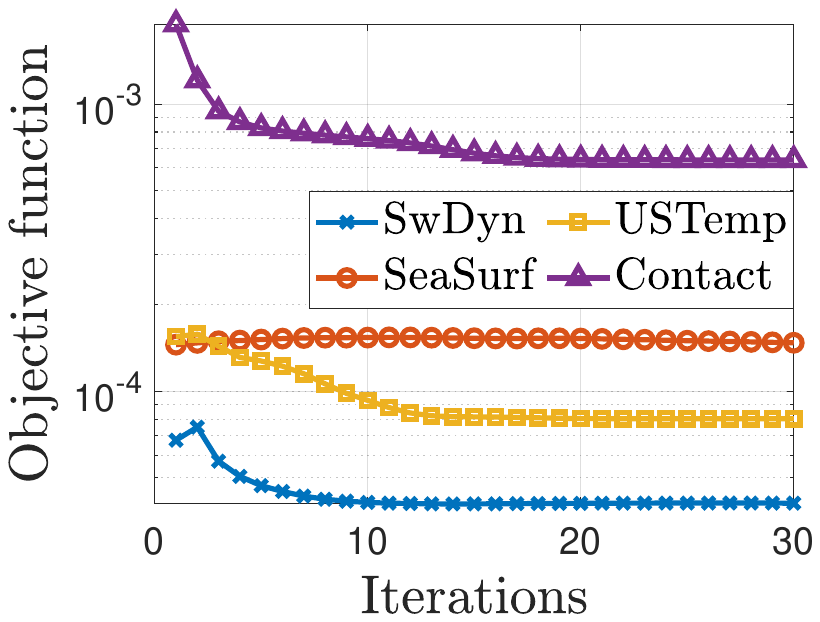}
\includegraphics[trim=100 250 100 300,width=0.23\textwidth]{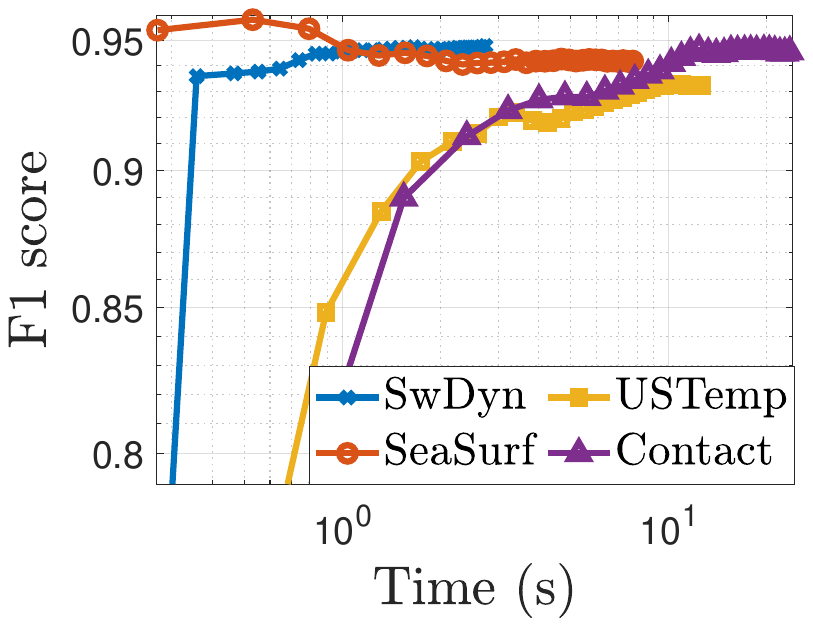}
\caption{Convergence of the DGD algorithm for different datasets with $R=10$ and $90\%$ of the temporal network as observed in terms of (a) the normalized objective function as the number of iterations increases and (b) the obtained $\fF1$ score together with the elapsed time for convergence.}
\label{F:convergence}
\vspace{-0.4cm}
\end{figure}
\vspace{0.05cm}
\par\noindent\textbf{Evaluation.} We evaluate the performance w.r.t. estimating the hidden dynamic network over the unobserved part of the true topology \(\uvbA\). To this end, we generate a binary mask \(\uvbM_{un} \in \{0,1\}^{N\times N\times T}\), whose nonzero entries correspond to unobserved entries in the original network. These entries can be associated with either edges or non-edges. Specifically, we construct \(\uvbM_{un}\) as the complement of a given mask \(\uvbM\), i.e., \([\uvbM_{un}]_{i,j,t} = 1 - M_{i,j,t}\). The mask \(\uvbM\) is designed to randomly mask both existing and non-existing edges with uniform probability across the tensor \(\uvbA\). This allows us to evaluate the estimation performance in regions of the network that were not observed.
We are interested in two types of error metrics:

\begin{enumerate}
    \item The relative error (RE), which is given as
    \begin{equation}
    \fRE={||\uvbM_{un}\circ(\hat{\uvbA}-\uvbA)||_F^2}/{||\uvbM_{un}\circ\uvbA||_F^2}
    \end{equation} 
    \item The \fF1 score, which is given as
    \begin{equation}
    \fF1 = {2\fPr\fRe}/{(\fPr+\fRe)},
\end{equation}
where $\fPr$ (precision) indicates the percentage of estimated edges that are edges of the unobserved dynamic network and $\fRe$ (recall) is the percentage of existing edges that were correctly estimated.
\end{enumerate}

\subsection{Initial DGD assessment}
In this section we investigate the performance of the proposed DGD method in terms of convergence, the role of the latent factors and its ability to recover the dynamic topology.
\vspace{0.05cm}
\par\noindent\textbf{Convergence.}
We assess the convergence of DGD by fixing the number of latent graphs \( R=10 \) and using $90\%$ of the temporal network topology in $\uvbA$ for all datasets. Figure \ref{F:convergence}.a illustrates the convergence behavior, while Figure \ref{F:convergence}.b shows the \(\fF1\) score as a function of elapsed time. Our approach converges within approximately 20 iterations across all datasets, except for USTemp. Figure \ref{F:convergence}.b further indicates that achieving convergence in the objective function correlates with the effective recovery of the ground-truth temporal network structure. The convergence is slower in larger networks.
\vspace{0.05cm}
\par\noindent\textbf{Effect of $R$ and observed topology.}
Next, we analyze the  proposed approach across two distinct setups:
\par
\begin{figure}[t]
\includegraphics[trim=100 250 100 300,width=0.23\textwidth]{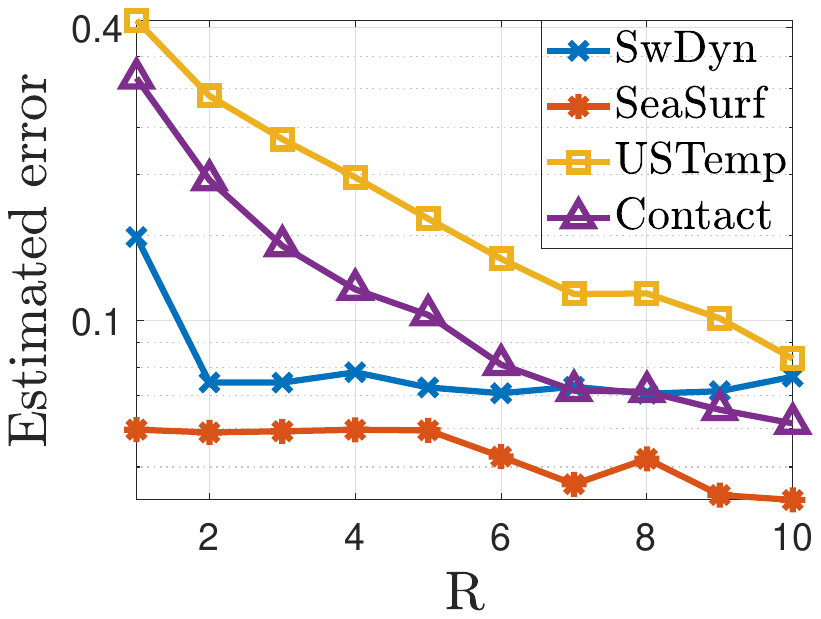}
\includegraphics[trim=100 250 100 300,width=0.23\textwidth]{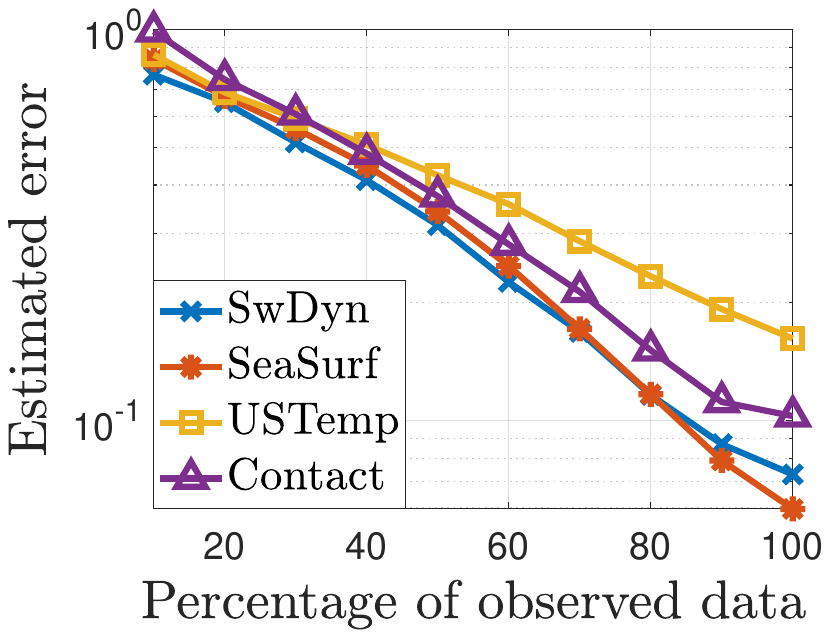}
\caption{Numerical evaluation of the DGD algorithm: (left) Relative error \fRE~w.r.t. the number of latent graphs $R$; (right) \fRE~w.r.t. the percentage of the observed temporal tensor $\uvbA$. } 
\label{Error vs r and obs}
\vspace{-0.4cm}
\end{figure}

\emph{Role of $R$}. Figure \ref{Error vs r and obs}.a illustrates the variation in $\fRE$ w.r.t. the number of latent graphs \(R\). We utilize $100\%$ of the graph signals and consider that $90\%$ of the entries of $\uvbA$ are observed. Increasing the number of latent graphs improves the approximation of the temporal network across all datasets. For the SwDyn dataset, a substantial improvement is observed when increasing \(R\) from 1 to 2, after which the error stabilizes. This matches the way the network was generated. It should be noted, however, that identifying the optimal \(R\) remains a non-trivial task. For a fixed $R$, the relative errors for different datasets can be attributed to their characteristics.
In the case of the USTemp dataset, the larger error stems from factors such as the high variation in the true topology $\uvbA$ and the less smooth signals over the $T=15$ time stamps (i.e., higher $\QV$ compared to the other datasets). This essentially indicates that networks with these characteristics are more challenging to reconstruct under the assumptions considered in our problem formulation, requiring a larger value of $R$ to better capture the temporal dependencies.

\par\emph{Role of observed data.} Figure \ref{Error vs r and obs}.b presents the \(\fRE\) as a function of the percentage of the observed temporal network for various datasets, using \(R=10\) and $100\%$ of the graph signals. The results show that reconstructing the original tensor from a limited portion of the observed network is more challenging, resulting in higher \(\fRE\). Consistent with previous findings, the quality of the estimation depends not only on the proportion of links observed but also on the characteristics of the setup, such as the structure of the network and the temporal variations in signal smoothness. This is evident from the higher estimation errors observed for the USTemp and Contact datasets.


\begin{figure*}[t]
\centering
{\includegraphics[trim=100 250 100 300,width=0.33\textwidth]{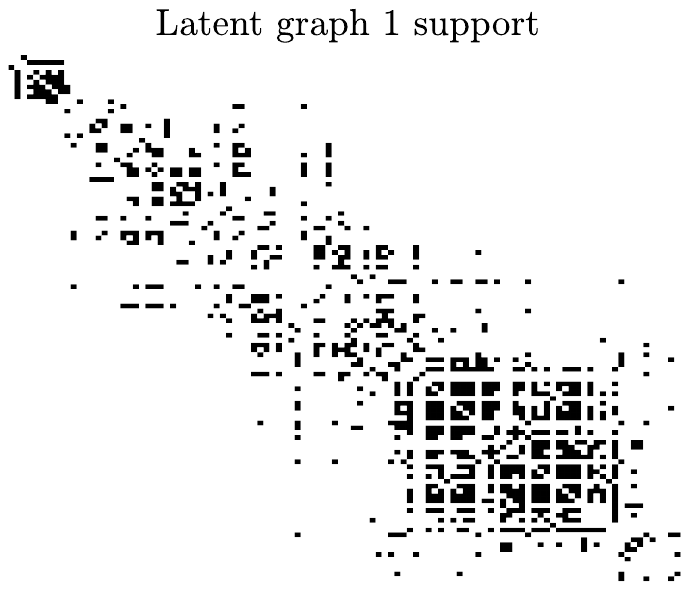}}%
\includegraphics[trim=100 250 100 300,width=0.33\textwidth]{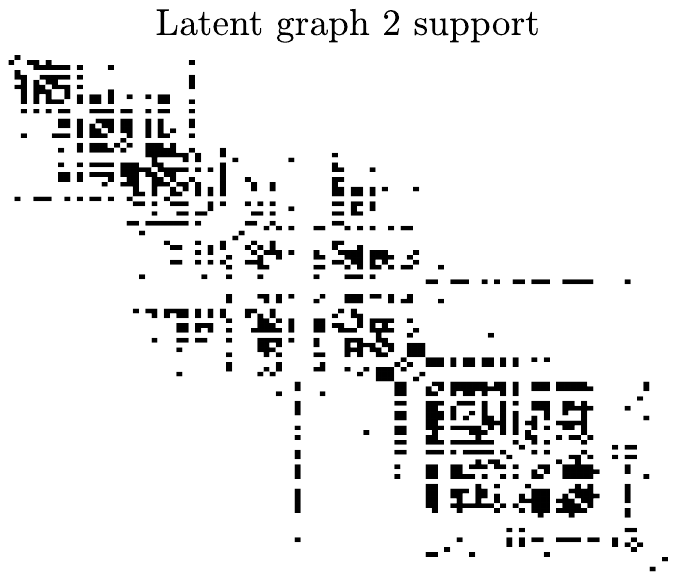}%
{\includegraphics[trim=100 250 100 300,width=0.33\textwidth]{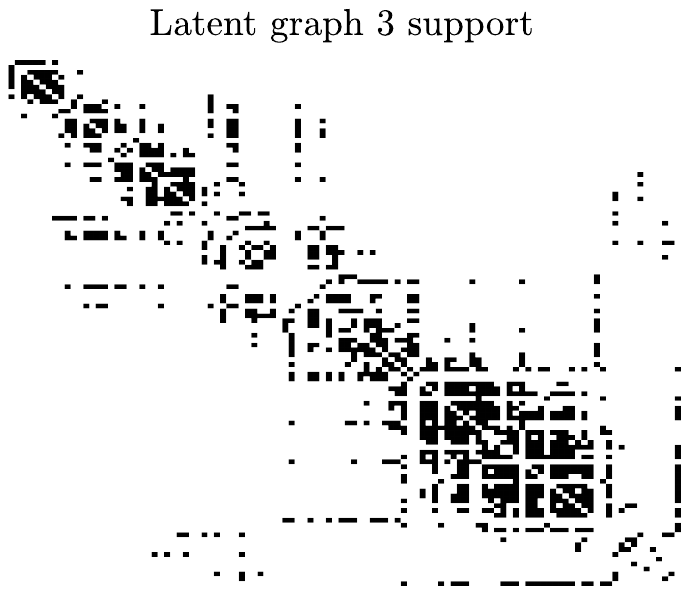}}%
\\
{\includegraphics[trim=100 250 100 300,,width=0.33\textwidth]{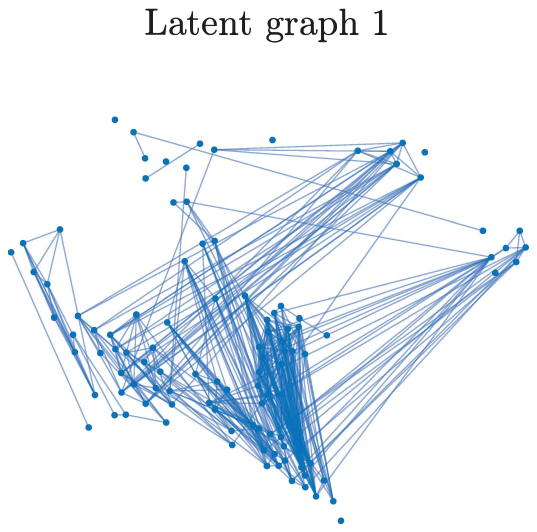}}%
\includegraphics[trim=100 250 100 300,width=0.33\textwidth]{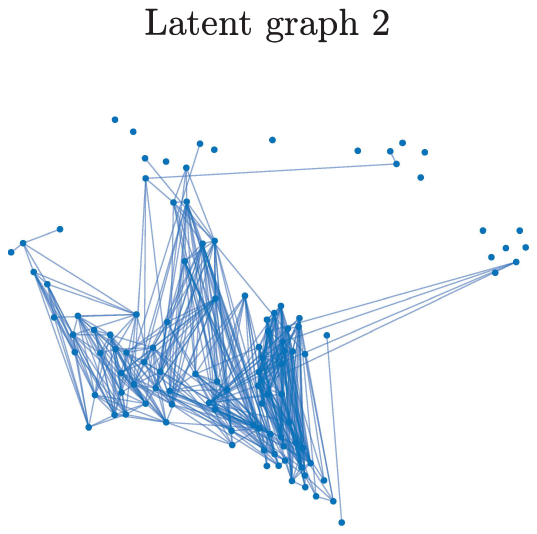}%
{\includegraphics[trim=100 250 100 300,width=0.33\textwidth]{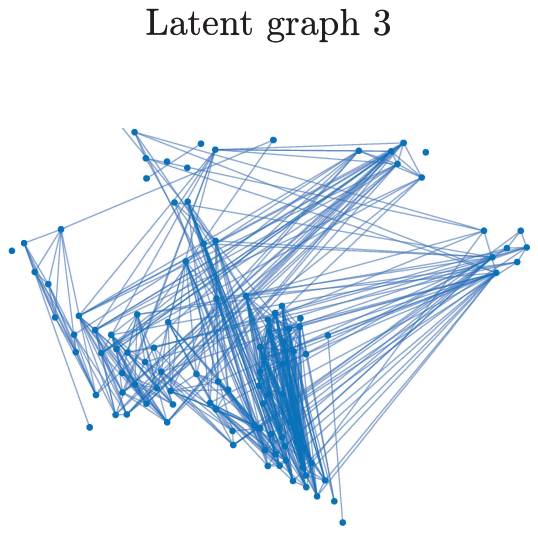}}
\vspace{-1cm}
\caption{Visualization of the components obtained from the DGD algorithm on USTemp dataset for $R=3$: (top row) the recovered latent adjacency matrices;  (bottom row)) shows the associated graph visualizations.}
\label{F:UStemp_graphs} 
\end{figure*}

\vspace{0.2cm}
\par
\begin{figure*}[t]
\centering
{\includegraphics[width=0.30\textwidth]{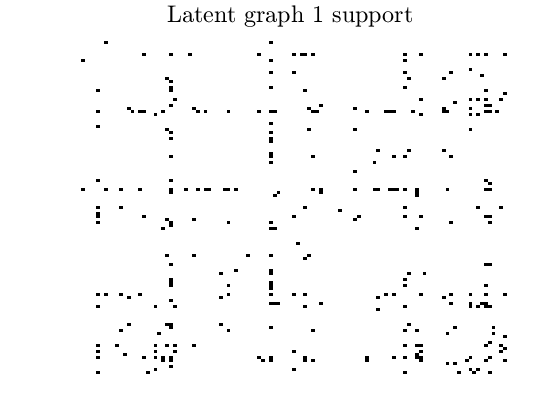}}%
\includegraphics[width=0.30\textwidth]{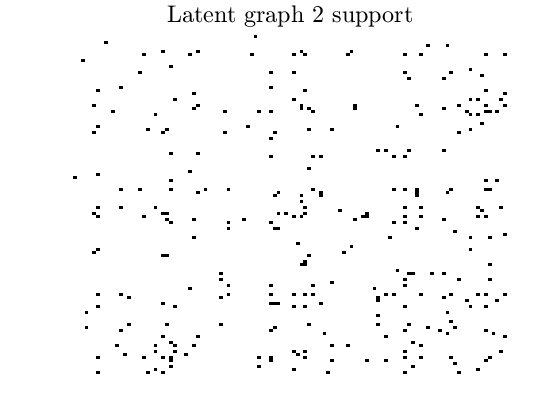}%
{\includegraphics[width=0.30\textwidth]{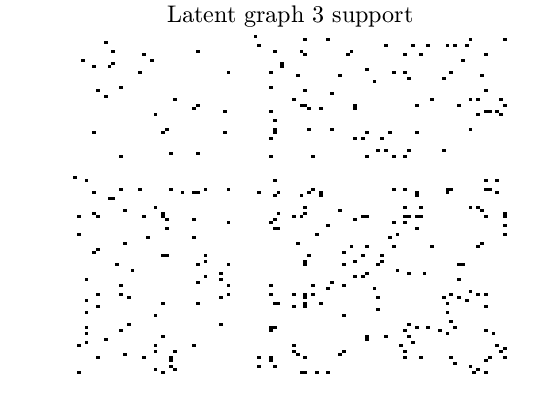}}%
\\
{\includegraphics[width=0.33\textwidth]{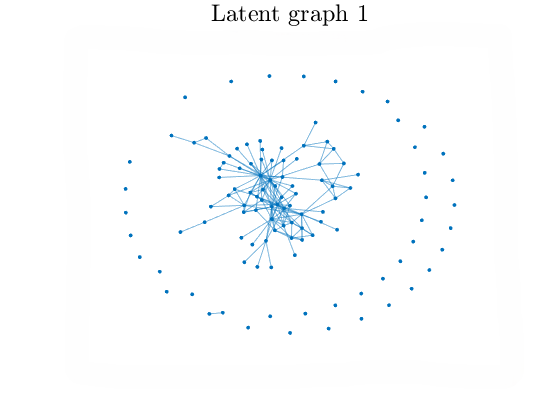}}%
\includegraphics[width=0.33\textwidth]{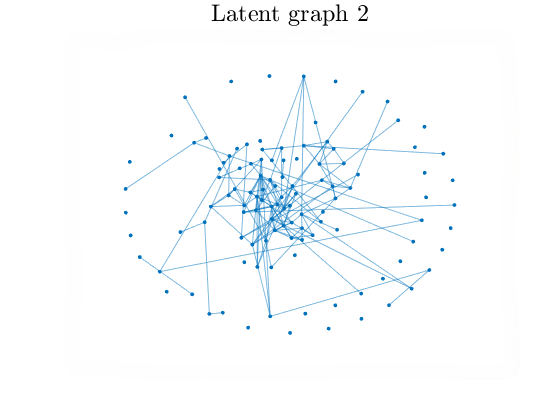}%
{\includegraphics[width=0.33\textwidth]{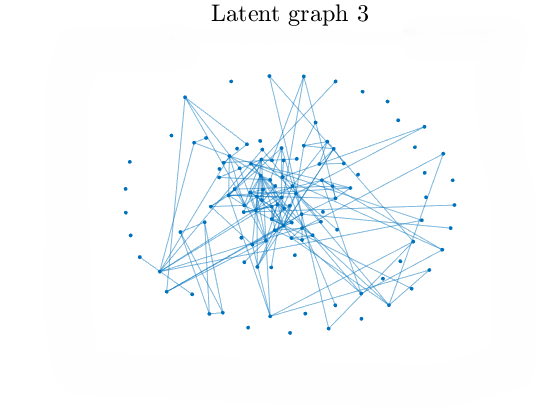}}
\vspace{-0.7cm}
\caption{Visualization of the components obtained from the DGD algorithm on Contact data for $R=3$: (top row) the recovered latent adjacency matrices;  (bottom row)) shows the associated graph visualizations.}
\label{F:contact_graphs} 
\end{figure*}

\begin{figure}
\includegraphics[trim=100 250 100 300,width=0.22\textwidth]{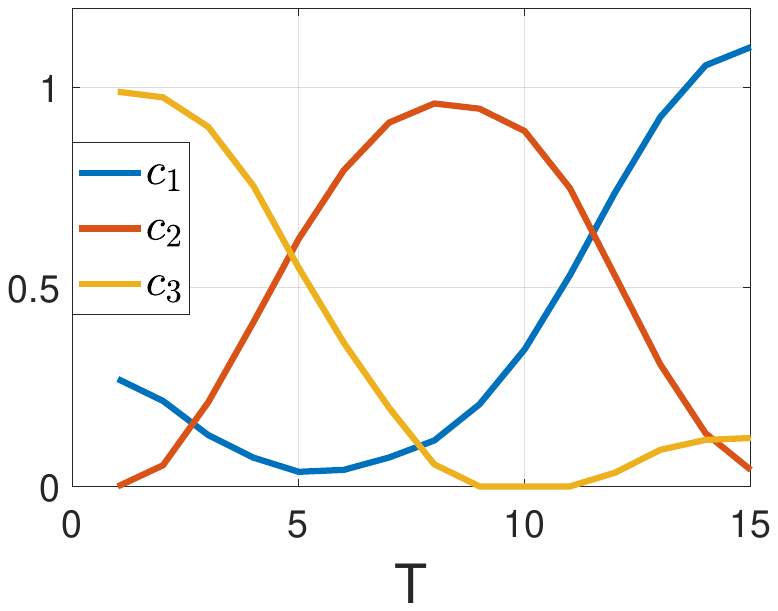}
\includegraphics[trim=100 250 100 300,width=0.22\textwidth]{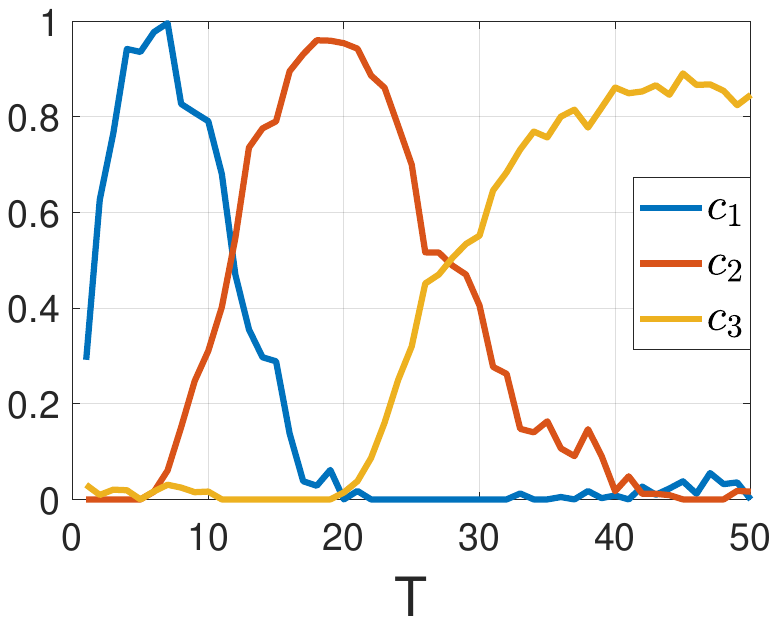}
\caption{Temporal signatures associated with each one of the latent adjacency matrices for (left) USTemp [cf. \ref{F:UStemp_graphs}] and (right) Contact data [cf. \ref{F:contact_graphs}].}
\label{F:Ustemp_signt}
\vspace{-0.7cm}
\end{figure}

\noindent\textbf{Recovered components.}
Next, we set $R=3$ and analyze the latent adjacency matrices \(\bbA_r\) and their corresponding temporal signatures \(\bbc_r\) for several datasets.
Figures \ref{F:UStemp_graphs} and \ref{F:contact_graphs} illustrate the three corresponding latent graphs. Each figure comprises two rows: the first row displays the support for each latent adjacency matrix, while the second row shows the corresponding graph representation. The latent graphs are individually and collectively significant both in terms of representing the true $\uvbA$ as well as reconstructing missing observations in $\uvbM_{un}\circ\uvbA$. We also see that each latent graph (and therefore its adjacency matrix) differs from the rest. They might share some edges but there are plenty of different edges as well. For USTemp, the first and third latent graphs have a similar structure which may be due to form sort of periodicity in the measurements.
Figure \ref{F:Ustemp_signt} illustrates the temporal signatures $\bbc_1,\bbc_2,$ and $\bbc_3$ associated with the latent graphs for the USTemp and Contact datasets shown in Figures 4 and 5. For USTemp, each signature is larger than the others over certain time intervals and changes smoothly over the $15$ time samples. A similar trend can be observed for the Contact dataset with slightly more fluctuations over time.
\vspace{0.2cm}

\begin{figure*}
\centering
{\includegraphics[trim=100 200 100 300,width=0.24\textwidth]{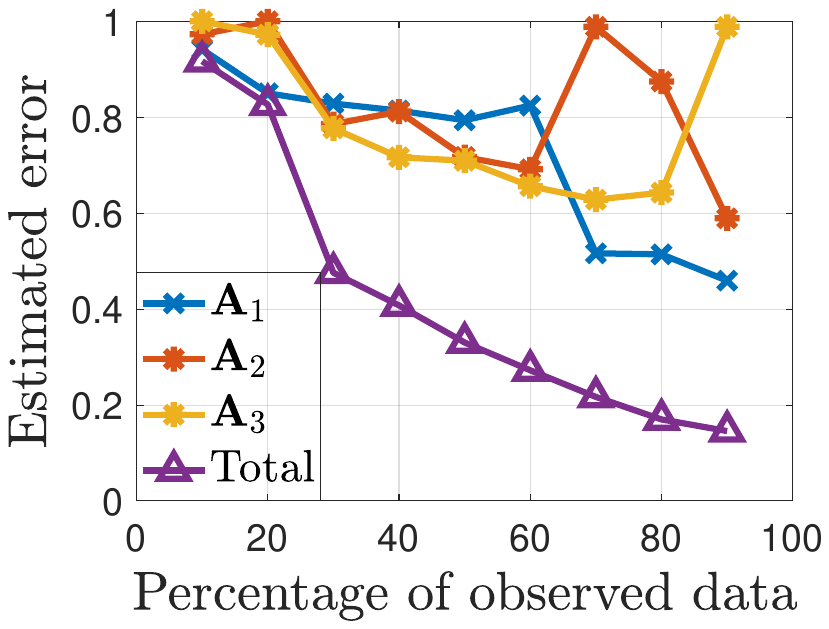}%
\includegraphics[trim=100 200 100 300,width=0.24\textwidth]{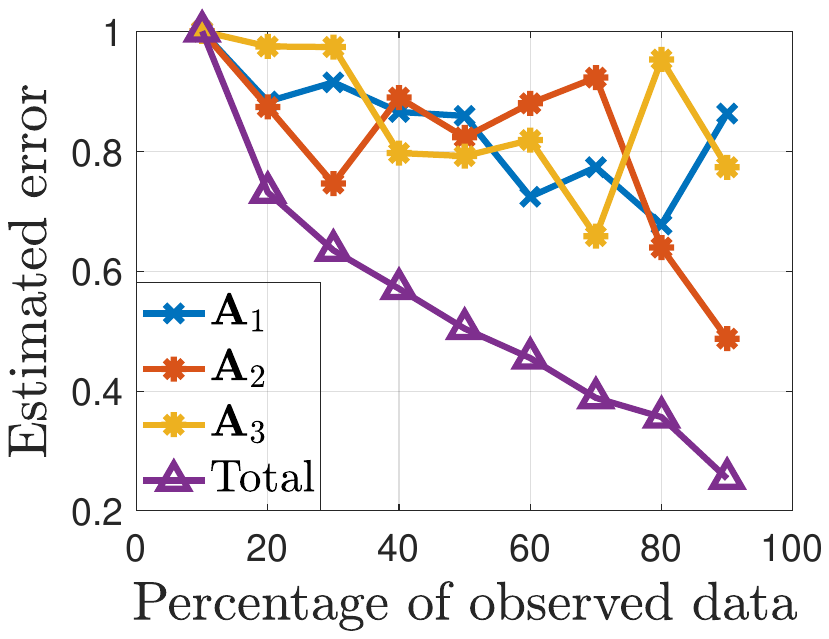}%
\includegraphics[trim=100 200 100 300,width=0.24\textwidth]{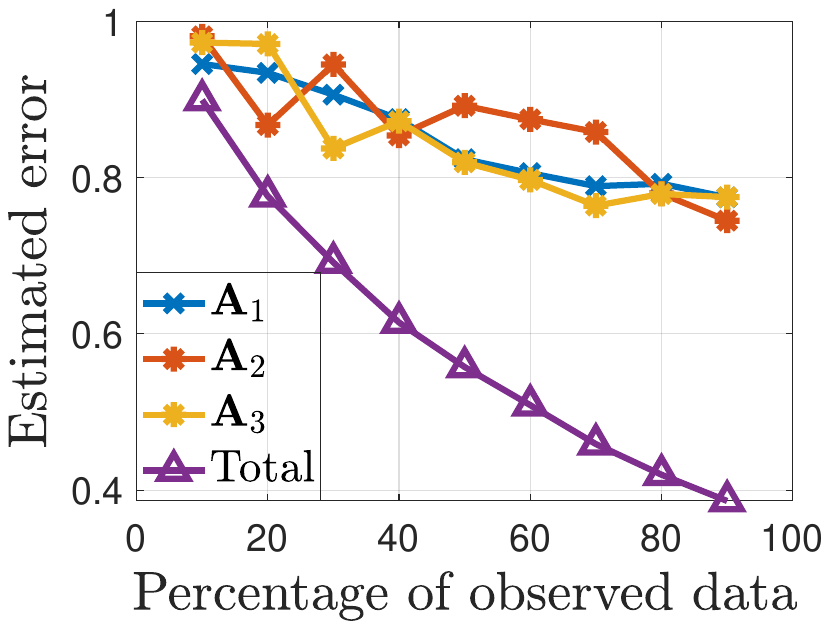}%
\includegraphics[trim=100 200 100 300,width=0.24\textwidth]{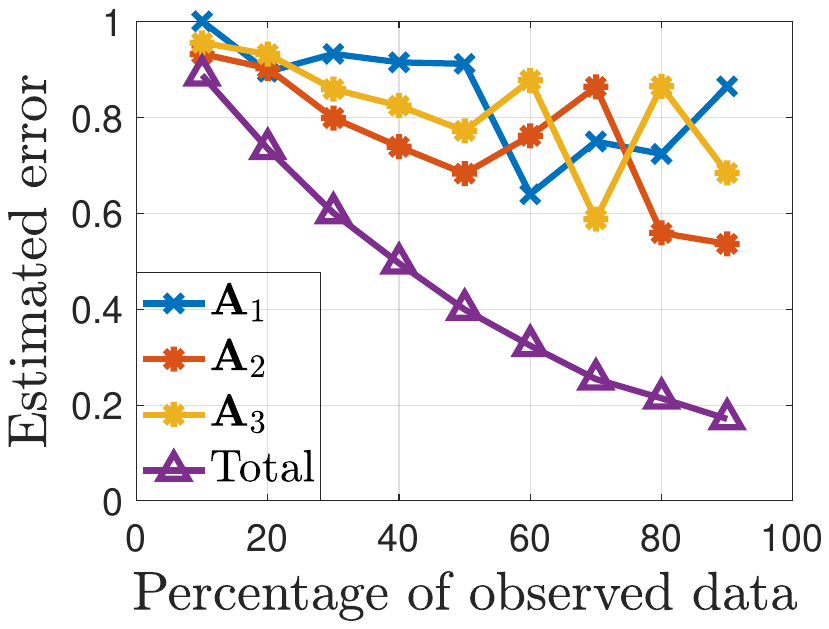}}%
\vspace{-0.6cm}
\caption{Reconstruction potential of $R=3$ individual latent graphs (blue, orange and yellow) for each dataset compared to that of the their model-based combination (purple). Here we show $\fRE$ for (left) SwSyn, (centre-left) SeaSurf, (centre-right) USTemp, and (right) Contact data.} 
\label{F:err_supp_reconstr}
\vspace{-0.5cm}
\end{figure*}

\par\noindent\textbf{Support recovery potential of recovered components.}
Next, we assess the contribution of individual estimated latent graphs towards the reconstruction of the temporal network. Specifically, we evaluate how accurately each latent graph captures the network's unobserved structure via the relative reconstruction error $\fRE$.
Figure \ref{F:err_supp_reconstr} shows the \(\fRE\) across the different datasets. For each dataset, we consider \(R=3\) and examine the reconstruction quality for each latent graph with \(\bbA_1\), \(\bbA_2\), \(\bbA_3\)—as well as the reconstruction achieved by combining all latent adjacency matrices. Individual graphs alone do not adequately capture the temporal network's structure, as evidenced by higher \(\fRE\) scores for each latent graph by itself. However, combining the graphs significantly improves the \(\fRE\) scores for all datasets. This demonstrates that while each latent graph contributes to explaining a portion of the temporal network $\uvbA$, no single latent graph is redundant. One reason for this is seen from the corresponding temporal signatures in Figure 5. Since each latent graph has a temporal signature that has high values over a slice of the temporal window, this graph itself would result in a lower error in recovering missing observations over this time period. And since the different temporal signatures peak at different times, the corresponding graphs recover different parts of $\uvbM_{un}\circ\uvbA$ with high accuracy. This can also be explained in part by the hyper-parameter $\beta$, which penalizes shared edges across the latent graphs. Higher values of $\beta$ can also cause edges to be separated across different $\bbA_r$s. Even if they have similar temporal signatures, they would still be able to reconstruct different aspects of $\uvbA$.
Overall, this experiment highlights the importance of considering all latent graphs collectively to achieve a comprehensive reconstruction of the temporal network. The corresponding plots for \fF1 are shown in Appendix \ref{J3 Appendix component-wise F1}.

\subsection{Comparison}
We assess the effectiveness of our approach in reconstructing the unobserved temporal network by comparing its recovery performance with that of other alternatives in various scenarios. Specifically, we compare DGD with the following approaches:
\begin{enumerate}
    \item \textbf{Unconstrained solution (UNC).} This is the unconstrained solution to Problem \eqref{Opt prob basic}, where we drop both the graph and temporal smoothness constraints on the latent adjacency matrices $\bbA_0$ and the temporal signatures $\bbC$, respectively. The solution is obtained by minimizing 
    \begin{equation}\label{E:unc}
\underset{\bbA_0,\bbC}{\text{minimize}}~||\bbA_{\textnormal{vec}}-\bbA_0\bbC^{\top}||_F^2
    \end{equation}where $\bbA_{\textnormal{vec}}$ and $\bbA_0$ are defined in \eqref{eq X_0} and \eqref{J3 Define A0}, respectively.
    \item\textbf{Canonical polyadic decomposition (CPD).} The CPD \cite{kolda2009tensor} approximates $\uvbA$ via a sum of three-dimensional vector outer products. We do not impose any graph constraints on the CPD, so the components may not be as interpretable like ours. 
\item \textbf{Block term decomposition (BTD).} The BTD approximates $\uvbA$ as a sum of outer products of matrices \cite{de2008decompositions}. Thus, we expect the BTD to represent more structure than the rank one terms in the CPD. We consider the $\{L,L,1\}$ type BTD, where each block term comprises the outer product of two matrices of rank $L$ and a vector of size $T$. We use the code in \footnote{\url{http://dimitri.nion.free.fr/Codes/Tensor_Decompositions.html}}.
    \item \textbf{No signal dynamic graph decomposition (NSDGD).} This is the proposed approach without incorporating the graph signals, i.e., $\delta=0$. This allows us to see the effect of including graph signals in the design.
    \item \textbf{Smooth graph learning (SGL).} For this baseline, we use the observed links, the observed graph signals and the smoothness prior to estimate the topology. Specifically, for each $t$, we estimate $[\uvbA]_{:,:,t}$, the topology at time $t$, using the signals in $\bbX_t$ and the graph-learning algorithm in \cite{kalofolias2016learn}.
    \item \textbf{SICA.} This approach, presented in \cite{hyvarinen2000independent}, approximates  tensor $\uvbA$ using a spatial independent component analysis. Thus, it outputs latent graphs that are independent in a statistical sense, without relying on graph-based constraints. 
\end{enumerate}
For a fair comparison between the different approaches, we ensure that CPD, BTD, NSDGD, and SICA have a similar number of degrees of freedom as DGD. This results in a higher number of terms in the CPD and BTD approaches, which may further complicate the ability to interpret the decomposition.
%
\begin{figure*}
\centering
    \includegraphics[trim=100 200 100 300,width=0.24\textwidth]{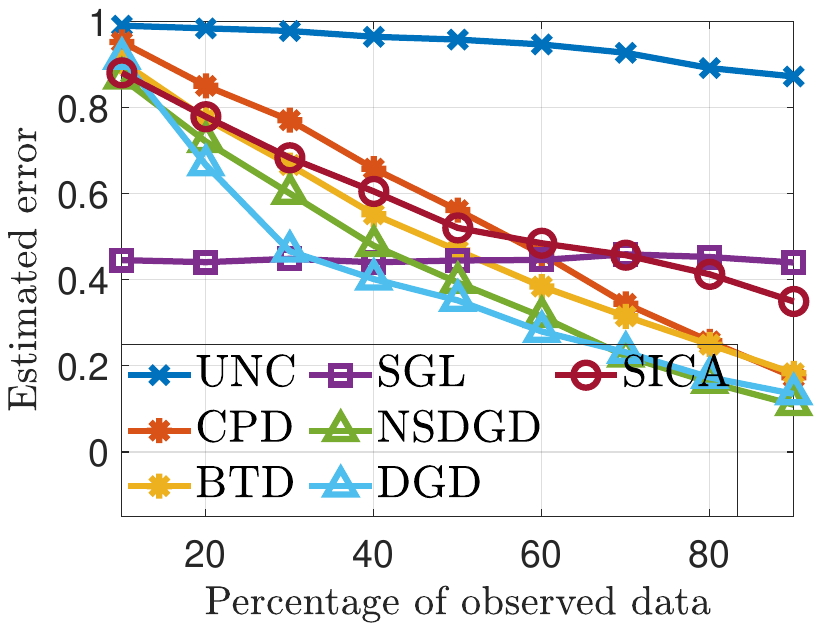}
    \includegraphics[trim=100 200 100 300,width=0.24\textwidth]{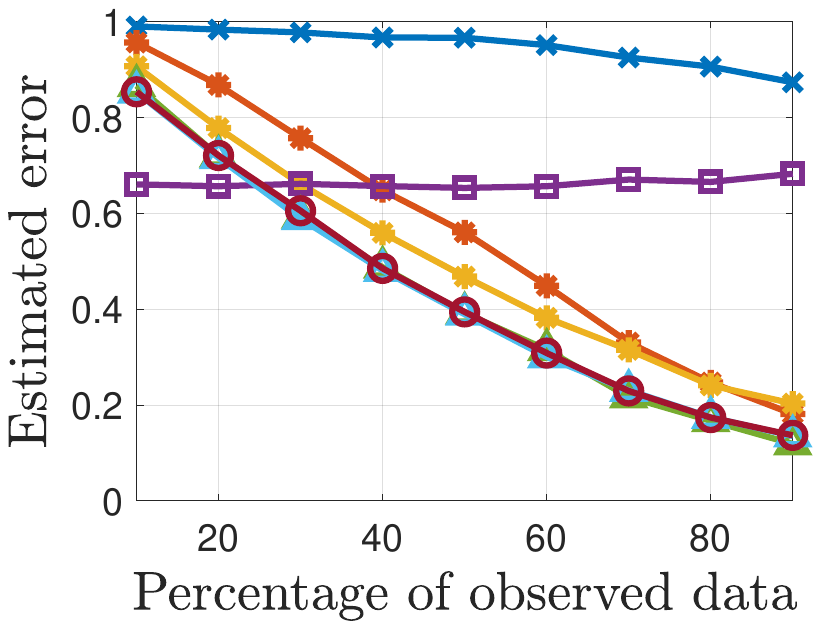}
    \includegraphics[trim=100 200 100 300,width=0.24\textwidth]{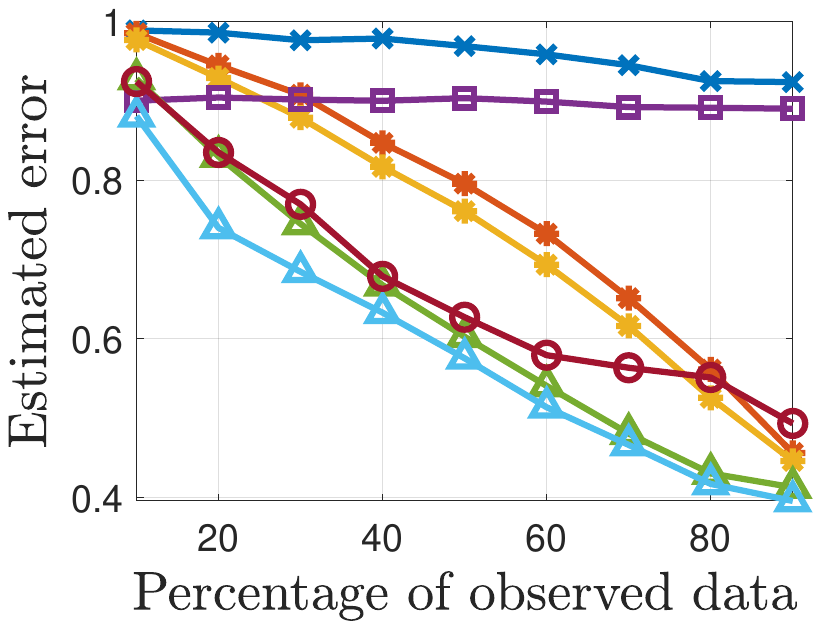}
    \includegraphics[trim=100 200 100 300,width=0.24\textwidth]{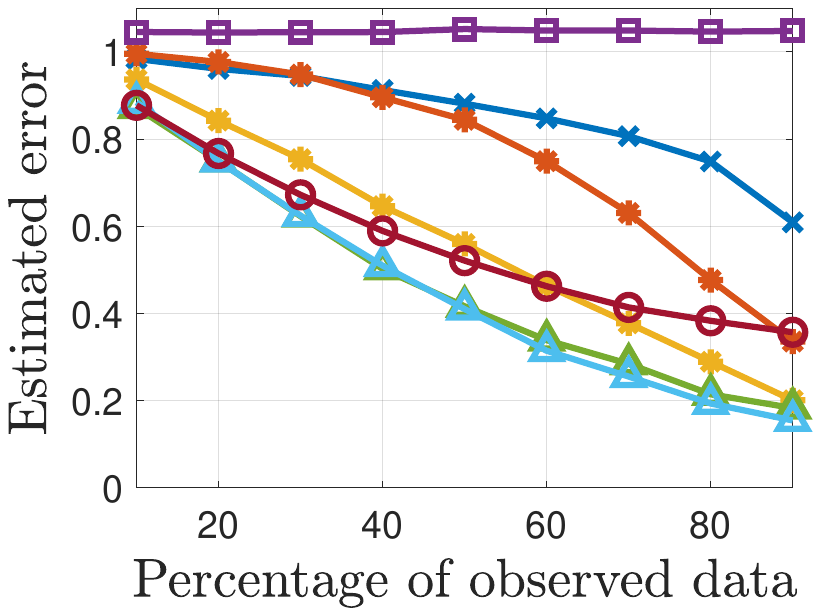}
    \vspace{-0.6cm}
    \caption{Relative error ($\fRE$) for different approaches as a function of the percentage of observed adjacency tensor $\uvbA$ for (left) SwDyn, (centre-left) SeaSurf, (centre-right) USTemp, and (right) Contact datasets, respectively. The legend in Figure 8 (left) applies to all the subfigures.} 
    \label{F:exp4_all_datasets}
    \vspace{-0.5cm}
\end{figure*}
\par Figure \ref{F:exp4_all_datasets} presents the $\fRE$ of each approach when varying the percentage of observed topology in the temporal network.
The results indicate that UNC tends to overfit the observed part of the network and fails to reconstruct the unobserved portions of the temporal network. DGD performs better than the low-rank tensor decomposition-based approaches (CPD and BTD) for all datasets for all percentages of observed topology. The gap in performance varies, from high (USTemp) to low (SwDyn). This corroborates our hypothesis that low-rank tensor decompositions are not very informative or representative of the structural evolution. BTD yields slightly better \(\fRE\) scores than CPD. Regarding the results associated with the SICA approach, we observe a slightly improved performance in terms of error compared to low-rank tensor decomposition methods. A notable outcome, as evidenced by the results, is that the SICA approach yields higher errors compared to NSDGD and DGS when the percentage of available data is large, while the opposite trend is observed with a smaller percentage of available data. This effect stems from the assumption in the SICA approach of estimating independent components, which may be overly restrictive for capturing the structure of the network when a substantial amount of information is available but becomes beneficial when data availability is limited. 

\par The DGD and NSDGD demonstrate superior performance compared to other methods. These approaches employ a more sophisticated temporal network structure, by incorporating the latent adjacency matrices and temporal signatures. Notably, DGD achieves the best performance, primarily due to its utilization of additional node signals. This additional information enhances the reconstruction accuracy, particularly in scenarios where a significant portion of the temporal network is unobserved. The gap between DGD and NSDGD narrows as we observe more of the structural evolution, showcasing the need to use the signals as a prior with limited observations. However, we note that SGL outperforms all approaches for SwDyn data when we observe $10$ to $40$ percent of $\uvbA$, i.e., just using the signals with signal smoothness prior recovers more of the missing structural evolution. This makes sense, as we generate smooth signals from the topology. This also suggests that a low percentage of observations hinders the recovery of hidden observations when combined with signals. For the other datasets, SGL struggles to recover the structural evolution.
\par In summary, the proposed approach shows advantages over existing methods by leveraging additional node signal information and employing it to improve the reconstruction error of the temporal network in scenarios with a large percentage of unobserved part of the temporal network.

\section{Conclusion}\label{J3 Section Conclusion}
We introduced a new approach to represent and estimate dynamic networks by modeling their structural evolution as weighted combinations of latent graphs, along with temporal signatures that modulate their importance over time. A key feature of our method is the integration of network observations with partial observations of spatiotemporal signals, assumed to be smooth on the dynamic network. By combining these data sources, we formulated an optimization problem whose solution yields the estimated latent graphs, temporal signatures, and dynamic network. We solved this problem via an alternating minimization scheme and showed that it converges to a block coordinate minimizer. Empirically, each recovered latent graph provides meaningful insights, contributes to representing the overall dynamics, and helps recover missing parts of the network. Numerical results further demonstrated that our method outperforms topology-agnostic low-rank tensor decompositions and approaches relying solely on graph signal smoothness. Future research includes developing an online extension of the decomposition, where the entire sequence of topologies is not available at once and must be updated over time. Another direction is to extend the framework to expanding graphs, accommodating the arrival of new nodes as the graph evolves.

\appendices
\section{Convergence Proofs}\label{J3 Appendix Convexity in A_rs and C and A}
\par\noindent\textbf{Strong Convexity of $f(\cdot)$ in $\bbA_r$.} We compute the Hessian and then show $\nabla^2_{\bbA_r}f\succ 0$. We start with the first-order derivatives of $f(\bbA_r)$ and the respective Hessians, which are 
\begin{alignat}{2}\label{Appendix derivative A_r}
\nabla_{\bbA_r}f(\bbA_r)\!&=\bbA_r(\sum_{t=1}^T[\bbC]_{t,r}^2\mathbf{1}_{N^2}^{\top}\bbm_t)
\!-[\sum_{t=1}^T\uvbA_{:,:,t}[\bbC]_{t,r}\mathbf{1}_{N^2}^{\top}\bbm_t] \nonumber
\\&+\big[\sum_{t=1}^T[\bbC]_{t,r}\mathbf{1}_{N^2}^{\top}\bbm_t\!\!\!\sum_{\bar{r}=1,\bar{r}\neq r}^R[\bbC]_{t,\bar{r}}\bbA_{\bar{r}}\big]+\!\gamma\mathbf1\mathbf1^{\top} +\\&+\delta{\bbXi_r^{\top}}\! \nonumber
+2\beta\sum_{k=1,k\neq r}^R\bbA_k\!.
\end{alignat} 
\begin{alignat}{2}\label{Appendix Hessian A_r}
\nabla^2_{\bbA_r}f(\bbA_r)&=\big(\sum_{t=1}^T[\bbC]_{t,r}^2\mathbf{1}_{N^2}^{\top}\bbm_t\big)\bbI_{N^2},
\end{alignat} 
with $\nabla^2_{\bbA_r}f(\bbA_r) \in \reals^{N^2\times N^2}$. 
From Assumption \ref{J3 Assumption 1}, we have strong convexity in $\bbA_r$.

\vspace{0.2cm}
\par\noindent\textbf{Strong Convexity of $f(\cdot)$ in $\bbC$.} For $\bbC$ we have
\begin{alignat}{2}\label{Appendix derivative C}
\nabla_{\bbC}f(\bbC)\!&=\!\bbF(\!-\!\bbA_{\textnormal{vec}}^{\top}\bbA_0\!+\!\bbC\bbA_0^{\top}\bbA)\! \nonumber \\ & +\!\delta(\bar{\bbZ}\bbA_0)\!+\!\mu\bbD^{\top}\bbD\bbC+\!\rho\bbC.
\end{alignat}
The Hessian w.r.t. $\bbC$ can similarly be derived as 
\begin{alignat}{2}\label{Hessian C}
& +\bbI_{R}\otimes(\bbD^{\top}\bbD+\rho\bbI_T)\text{vec}(\difd\bbC), \nonumber \\
\nabla^2_{\bbC}f(\bbC)&=\big(\bbA_0^{\top}\bbA_0\otimes\bbF+\bbI_{R}\otimes\bbD^{\top}\bbD+\rho\bbI_{RT}\big),
\end{alignat} 
where $\nabla^2_{\bbC}f(\bbC) \in \reals^{TR\times TR}$. 
The matrix $\big(\bbA_0^{\top}\bbA_0\otimes\bbF+\bbI_{R}\otimes\bbD^{\top}\bbD+\rho\bbI_{RT}\big)$ is composed of thee components. First, $\bbA_0^{\top}\bbA_0\otimes\bbF$ is a positive semi-definite matrix as both the matrices $\bbA_0^{\top}\bbA_0$ and $\bbF$ (which is a diagonal matrix with positive diagonal elements) are positive semi-definite. By the same argument, $\bbI_{R}\otimes\bbD^{\top}\bbD$ is positive semi-definite. Since $\rho>0$, we have $\bbA_0^{\top}\bbA_0\otimes\bbF$ is a positive semi-definite matrix as both the matrices $\bbA_0^{\top}\bbA_0$ as positive definite, which means $\nabla^2_{\bbC}f\geq \rho\bbI_{RT}$, establishing strong convexity in $\bbC$.\qed
\vspace{0.2cm}
\section{Proof of Proposition}\label{app:sectionB}
We prove the convergence of our alternating approach to a stationary point. Here, we denote $\bbA_{-r}$ as the set of all $\bbA_i$s with $i\neq r$.
To show that we converge to a block coordinate minimum, we must satisfy the requirements mentioned in Theorem 2.3, Lemma 2.2, and Assumptions 1 and 2 in \cite{xu2013block}. These are as follows.
\begin{itemize}
    \item $f(\bbA_1,\ldots,\bbA_R,\bbC)$ is block-multiconvex, i.e., it is convex in each $\bbA_r$ and $\bbC$, with the other block elements being fixed. This can be seen from Appendix \ref{J3 Appendix Convexity in A_rs and C and A}. In fact, we have strong convexity for each block element.
    \item Block multiconvexity feasible sets. That is, the sets corresponding to the constraints of each block element should be convex, given the others are fixed. We have $\sum_{r=1}^R\bbA_r\bbPhi_r-\zeta\mathbf{1}_{N\times T}\geq\mathbf{0}_{N\times T}$ which is block multi-convex, i.e., for fixed $\bbA_{-r}$ and $\bbC$, the set satisfying $\bbA_r\bbPhi_r+\bbGamma_r\geq\mathbf{0}$ is convex in $\bbA_r$. The same holds for $\bbC$ with the constraint set for $\bbA[\bbC^{\top}\otimes\mathbf{1}_N]\geq \zeta\mathbf{1}_{N\times T}$ being convex in $\bbC$ for fixed $\bbA_r$s.
    \item The minimum of $f()$ w.r.t. $\bbA_r$ exists and is finite. This is verified because for all $\bbA_r$ in the feasible set, the loss function $f()$ is nonnegative and bounded. Similarly, the minimum of $f()$ w.r.t. $\bbC$ is also finite, following the same reasoning, i.e., element-wise positivity.
    \item The set map w.r.t. each $\bbA_r$ changes continuously [cf. \cite{xu2013block}]. The constraint $\bbA_r\bbPhi_r+\bbGamma_r\geq\mathbf{0}$, which changes for every update, is continuous in each $\bbA_r$. The constraint set $\bbA[\bbC^{\top}\otimes\mathbf{1}_N]\geq \zeta\mathbf{1}_{N\times T}$ changes in a continuous manner for each update in $\bbC$, given the changes in $\bbA_r$. This is easily verified, given our alternating update scheme.
\end{itemize}
Therefore, the proposed alternating approach converges to a block coordinate minimizer in the latent adjacency matrices $\bbA_r$s and their temporal signatures in $\bbC$.\qed
\section{Gradients for ADMM updates}\label{J3 Appendix ADMM Updates}
The following is the gradient expression used in \eqref{ADMM_A}:
\begin{alignat}{2}
\nabla_{\bbA_r}L(\bbA_r)&=\bbA_r(\sum_{t=1}^T[\bbC]_{t,r}^2\bbone^{\top}\bbm_t)+\big[\sum_{t=1}^T[\bbC]_{t,r}\bbone^{\top}\bbm_t \nonumber \\
&\times\!\!\!\!\sum_{\bar{r}=1,\bar{r}\neq r}^R\!\!\!\![\bbC]_{t,\bar{r}}\bbA_{\bar{r}}\big]-\big[\sum_{t=1}^T\uvbA_{:,:,t}[\bbC]_{t,r}\tr(\diag(\bbM_t))\big]\nonumber \\
&+\delta{\bbXi_r^{\!\top}}\!+\gamma\mathbf1\mathbf1^{\!\top}\!+2\beta\!\!\!\!\sum_{k=1,k\neq r}^R \!\!\!\!\!\bbA_k+\eta\bbA_r+\bbLambda_r^{\!\top}\bbPhi_r^{\!\top}\! \nonumber \\
& +\lambda(\bbA_r\bbPhi_r\!+\!\bbGamma_r\!-\!\bbP)\bbPhi_r^{\top}.
\end{alignat}
The gradient of the augmented Lagrangian w.r.t. $\bbC$ as used in \eqref{LagrangianC} is
\begin{alignat}{2}
    \nabla_{\bbC}L(\bbC)&=\bbF(-\bbA_{\text{vec}}^{\top}\bbA_0+\bbC\bbA_0^{\top}\bbA_0)+\delta(\bar{\bbZ}\bbA_0)+\mu\bbD^{\top}\bbD\bbC \nonumber \\ 
    &+\rho\bbC+\bbLambda^{\top}\Upsilon+\lambda_c(\bbC\Upsilon^{\top}-\zeta\mathbf1_{T,N}-\bbQ)\Upsilon,
\end{alignat}
where $\bbF=\diag(\bbf)=\diag(\mathbf{1}_{N^2}^{\top}\bbM_0)$, i.e., the diagonal elements of $\bbF$ measure the number of active observations at that time.
\section{Component-wise $\fF1$ Scores}\label{J3 Appendix component-wise F1}
The following are the \fF1 scores corresponding to the $\fRE$ scores as shown in Figure \ref{F:err_supp_reconstr} in Section \ref{J3 Section Experimental}.
\begin{figure}[ht]
\centering
\includegraphics[trim=100 200 100 300,width=0.24\textwidth]{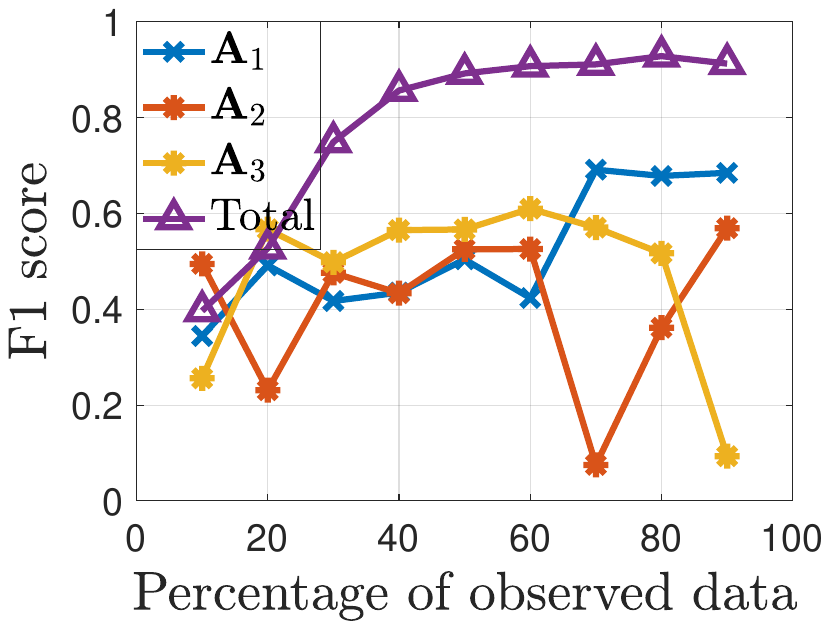}
\includegraphics[trim=100 200 100 300,width=0.24\textwidth]{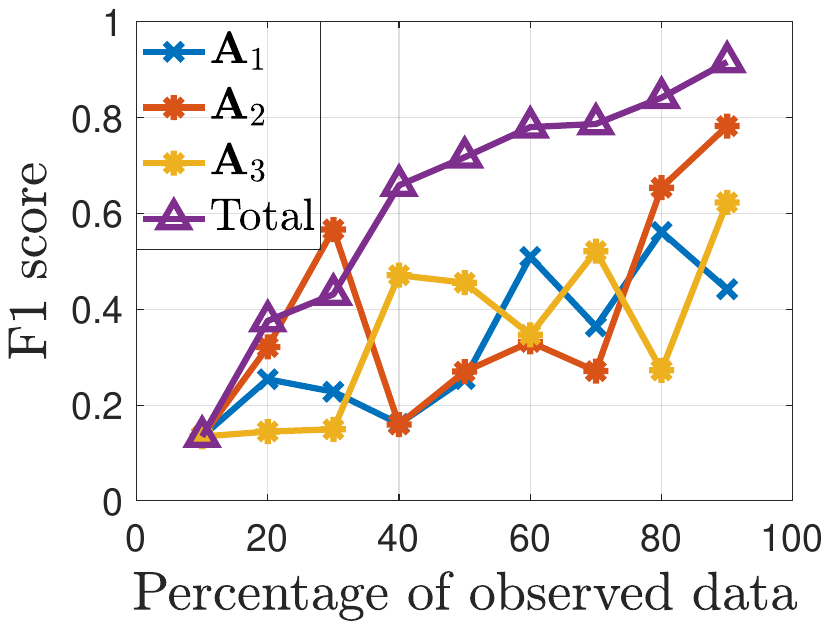}\\
\includegraphics[trim=100 200 100 300,width=0.24\textwidth]{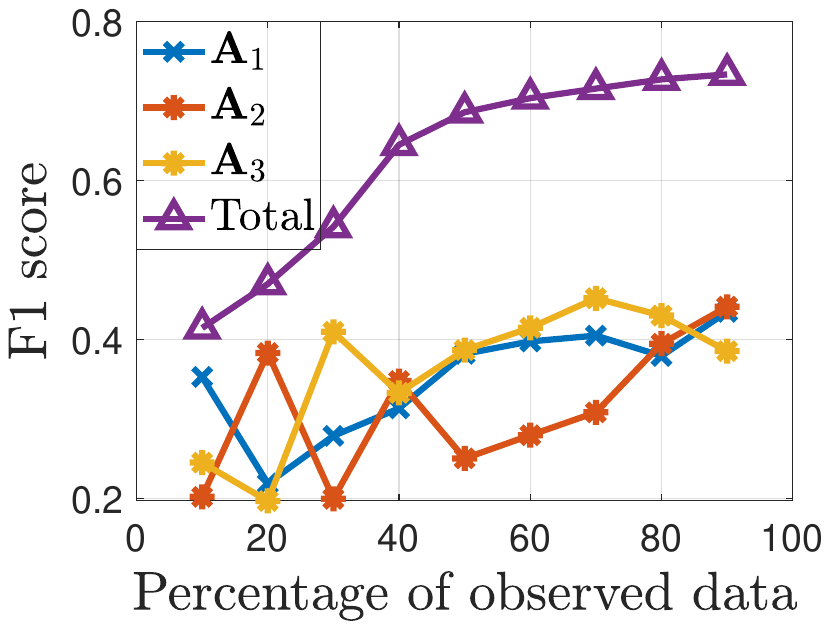}
\includegraphics[trim=100 200 100 300,width=0.24\textwidth]{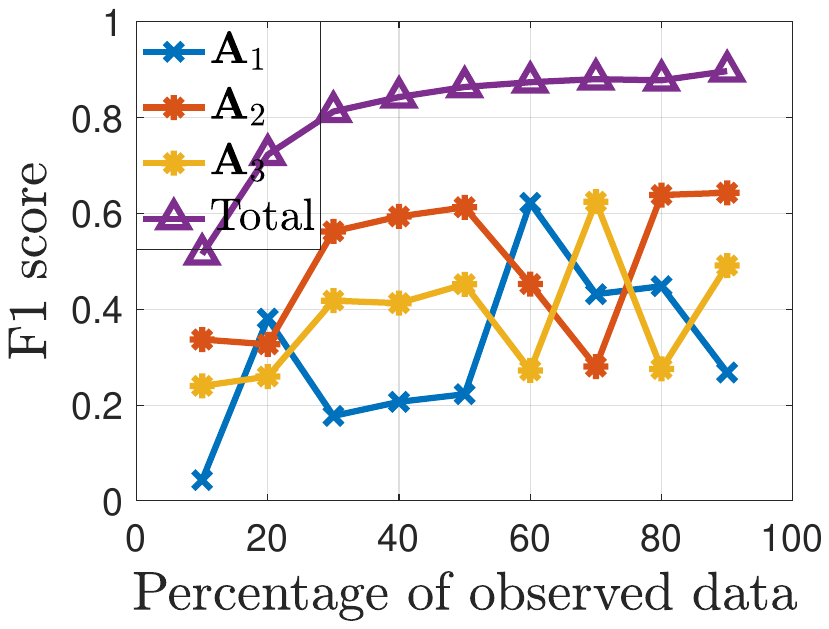}
\vspace{-1cm}
\caption{\fF1 of each of the $R=3$ latent graphs (blue, orange and yellow) for each dataset compared to that of the their model-based combination (purple)} 
\vspace{-0.5cm}
\end{figure}
\bibliographystyle{IEEEbib}
\bibliography{myIEEEabrv,strings}
\end{document}